\documentclass[letterpaper,12pt]{article}
\usepackage[usenames,dvipsnames]{xcolor}
\usepackage{mathtools} 
\usepackage{jheppub2} 
\usepackage[T1]{fontenc}
\usepackage{natbib}
\usepackage{graphicx}
\usepackage{hyperref}
\usepackage{amsmath}
\usepackage{amsfonts}
\usepackage{amssymb}
\usepackage{enumitem}
\usepackage{tocvsec2}
\usepackage{slashed}
\usepackage{comment}

\newcommand\beq{\begin{eqnarray}}
\newcommand\eeq{\end{eqnarray}}

\newcommand{\wtC}{{\widetilde C}}

\newcommand{\half}{\frac{1}{2}}

\newcommand\eqn[1]{\label{eq:#1}} 
\newcommand\eq[1]{eq. (\ref{eq:#1})}

\newcommand{\bfx}{{\mathbf x}}

\newcommand{\bfr}{{\mathbf r}}

\newcommand{\bfp}{{\mathbf p}}
\newcommand{\bfq}{{\mathbf q}}

\newcommand{\CA}{{\cal A}}

\newcommand{\CD}{{\cal D}}

\newcommand{\CI}{{\cal I}}
\newcommand{\CJ}{{\cal J}}

\newcommand{\CL}{{\cal L}}

\newcommand{\Tr}{{\rm Tr\,}}

\newcommand\vev[1]{\langle #1 \rangle}

\newcommand\ket[1]{| #1 \rangle}

\newcommand\expect[3]{\langle #1|#2|#3\rangle}
\newcommand{\mybar}[1]%
        {\kern 0.6pt\overline{\kern -0.6pt#1\kern -0.6pt}\kern 0.6pt}

\def\half{\tfrac{1}{2}}

\hypersetup{
    colorlinks=true,   
    linkcolor=Cerulean,     
    citecolor=OrangeRed,    
    filecolor=magenta, 
    urlcolor=blue      
}

\makeatletter
\def\l@subsubsection#1#2{}
\makeatother    

\advance\footnotesep 1.5pt

\begin{document}

\preprint{INT-PUB-22-013}
\title{Generalized Hall currents in topological insulators and superconductors}
\author[1]{David B. Kaplan,}
\emailAdd{dbkaplan@uw.edu}
\affiliation[1]{Institute for Nuclear Theory, University of Washington, Seattle WA 98195}
\author[2]{Srimoyee Sen,}
\emailAdd{srimoyee08@gmail.com}
\affiliation[2]{Department of Physics and Astronomy,  Iowa State University, Ames IA 50011}

\abstract{
We generalize the idea of the quantized  Hall current  to count gapless edge states in topological materials, applying equally well to theories in different dimensions, with or without continuous symmetries in the bulk or chiral anomalies on the boundaries. This current is related to the  index of the Euclidean fermion operator and can be calculated via one-loop Feynman diagrams.  Quantization of the current  is shown to be governed by topology in phase space, and the procedure can be applied to topological classes governed by either ${\mathbb Z}$ or ${\mathbb Z}_2$ invariants. We analyze several explicit examples of free fermions in relativistic field theories. We speculate that it may be possible to extend the technique to interacting theories as well, such as the interesting cases where  interactions   gap the edge states.}

\maketitle

\maxtocdepth{subsection} 

\section{Introduction}
\label{sec1}
Topological insulators and superconductors, often well described by free fermion theories \cite{Ludwig_2015, PhysRevB.85.085103, Ryu_2010, Kitaev:2009mg, PhysRevB.78.195125}, generally have gapless fermion excitations confined to the boundary between two different topological phases. For $d+1$ dimensional defects, where $d$ is odd, these massless fermions can be Weyl fermions. In that case the boundary theory has a chiral anomaly, and the currents for classical axial symmetries can have nonzero divergence in the presence of background gauge fields. These currents are conserved in the higher dimensional bulk theory, and the violation of axial charge in the boundary theory can be understood as the result of a fermion number current flowing in from the bulk, as in the Integer Quantum Hall Effect \cite{Callan:1984sa}. The argument can be turned around: the inflow of current from the bulk indicates a boundary theory with a chiral anomaly, which in turn requires massless particles to exist at the boundary. 

It is appealing that one can deduce the existence of gapless edge states from the inflow of current from infinity, but  it is not general. For example,  gapless edge states exist for superconducting systems for which there is no conserved fermion number current, as well as for boundaries with even spatial dimension $d$  where there are no chiral anomalies.   In a recent paper we showed, however, that in all of these cases one can generalize the idea of Hall currents and detect the existence  gapless edge states by this current's divergence \cite{kaplan2021index}.   That theory makes use of  the  index of the fermion operator in the Euclidian action  in the presence of external ``diagnostic fields''.  A very simple procedure for calculating the current flow  in terms of a one-loop Feynman diagram allows one to detect the existence of gapless edge states.  Furthermore, this construction makes the topological protection of the gapless modes manifest.  In the present paper we work through the examples given in Ref.~\cite{kaplan2021index} in greater detail, including the construction of the generalized Hall current,  the topological origins of its nonzero divergence, and the role of regularization.  We also extend the previous work with a discussion of the effect of interactions.  Our discussion throughout is in the context of relativistic quantum field theory.

 \section{General framework}
 \label{sec:2}
Our starting point is a relativistic  quantum field theory of free fermions in infinite $d+1$ dimensional Minkowski spacetime.  The action is   $S_\text{M}=\int d^{d+1}x\,\, \bar{\psi}\CD_\text{M}\psi$, where the fermion operator $\CD_\text{M}$ can include a position dependent mass term, {\it e.g.} one that exhibits defects like a domain wall or a vortex singularity, which will serve as proxies for a boundary. If the fermion operator $\CD_\text{M}$ hosts massless fermion states trapped in these defects, and the defects are space and time translation invariant in the co-dimensions, then the spectrum will include a state with zero momentum that is constant in the space-time coordinates of the defect, and is localized on the defect in the transverse direction(s). We wish to have a simple way to detect the existence of such states which makes manifest the underlying topology.  The approach we use is to note that when $\CD_\text{M}$ is analytically continued to Euclidian time,   the Euclidian fermion operator $\CD$ is an elliptic operator and the  propagator  $1/\CD$ will have poles corresponding to gapless Minkowski edge states with zero momentum.  We detect these states by computing the index of $\CD$ --- the number of zeromodes of $\CD$ minus that of $\CD^\dagger$. One obstacle to this program is that the index only counts normalizable zeromodes, while the states we are discussing are constant in the noncompact boundary dimensions; another is that $\CD^\dagger$ also exhibits poles corresponding to these states.  Both can be circumvented by adding ``diagnostic'' background fields that localize these states within the boundary without introducing a gap.  This can be easily done when the diagnostic fields possess nontrivial spatial topology; an interesting feature of this approach is that the diagnostic fields are typically not fields that could have been introduced into the original Minkowski theory, as we will explain below.  In the presence of these diagnostic fields,  solutions to $\CD \psi=0$ can be localized, while those to $\CD^\dagger\psi=0$ are delocalized, resulting in a nonzero index for  $\CD$.  This effect is seen with the toy operator $\CD= \partial_x + \epsilon(x)$ (where $\epsilon(x) \equiv x/|x|$ plays the role of the diagnostic field) which  has a localized solution  to $\CD\psi=0$  while the conjugate operator $\CD^\dagger=- \partial_x + \epsilon(x)$ does not. The index then indicates the presence of a massless edge state in the Minkowski theory, provided that it persists in the limit that the energy density  in the diagnostic fields  tends to zero.

 Computing the index then simply requires computing the divergence of an in-flowing current by means of a 1-loop diagram, which generalizes the in-flow picture from the quantum Hall system in any dimensions and whether the system has a conserved current or not.   Furthermore, the index  is shown to be the product of the winding number of the diagnostic fields in coordinate space times the winding number of the fermion dispersion relation  in momentum space, making manifest the topological nature of the gapless modes.    We will see that the Euclidian momentum space topology is simpler than the topological invariants of the Minkowski systems would suggest: in each case we examine it is governed by the homotopy group $\pi_n(S^n)$, where $n$ is the number of spacetime dimensions.  The index we compute reflects the correct invariant  of the Minkowski system, ${\mathbb Z}_2$ for example,  due to the interplay of momentum space and coordinate space topology in its definition.

Our procedure is (i) start with a Minkowski theory of interest; (ii) analytically continue to Euclidian space time; (iii) introduce diagnostic fields to localize edge states; (iv) compute the index of the Euclidian fermion operator in the limit of vanishing diagnostic fields, which involves computing the inflow of a generalized Hall current.  In the cases where fermion number is not conserved, relativistic systems analogous to topological superconductors, the Euclidian action takes the form $S_\text{E} = \frac{1}{2}\int \psi^T C \CD \psi$, where $C$ is the charge conjugation matrix. In these cases at step (ii)  we consider instead the Dirac action $S_E = \int \bar\psi \CD\psi$ and proceed from there, since solutions to $\CD\psi=0$ are also solutions to $C \CD\psi=0$. The Dirac theory allows for the addition of diagnostic fields not possible in the original theory, such as a $U(1)$ gauge field coupled to fermion number, or fields in $\CD$ which appear symmetrically in $C\CD$, and hence would not couple to the fermions in 
$\psi^T C \CD \psi$.

The procedure for computing the index of an elliptic operator $\CD$ is to first define
\beq
\CI(M)=\Tr\left(\frac{M^2}{\CD^{\dagger}\CD+M^2}-\frac{M^2}{\CD\CD^{\dagger}+M^2}\right)
\eqn{inddef}
\eeq
with
\beq
{\rm ind}(\CD) = \CI(0)\equiv \lim_{M\to 0} \CI(M)\ .
\eeq
The function $\CI(M)$ can be computed by means of Feynman diagrams by first defining  
\beq
K=\begin{pmatrix}
0 && -\CD^{\dagger}\\
\CD && 0
\end{pmatrix},\,\qquad
\Gamma_{\chi}=\begin{pmatrix}
1 && 0\\
0 && -1
\end{pmatrix}\ ,
\eqn{kdef}\eeq
in terms of which $\CI(M)$ is expressed as
\beq
\CI(M)=\Tr\left(\Gamma_{\chi}\frac{M}{K+M}\right).
\eqn{im}
\eeq
The matrix $K$ and $\Gamma$ have twice the dimension of $\CD$. The operator $1/(K+M)$ looks like a fermion propagator in a theory whose action is 
\beq
S_K =\int d^{d+1}x\, \mybar \Psi (K +M)\Psi,
\label{act}
\eeq
where $\Psi$ has twice the number of components as the original $\psi$ field off our Minkowski theory, but the integration is still over $d+1$ spacetime coordinates. 
In this extended theory, the quantity $\CI(M)$ in \eq{im} can be expressed as the matrix element of the pseudoscalar density, 
\beq
\CI(M)=-M\int d^{d+1}x\,\langle\bar{\Psi}\Gamma_{\chi}\Psi \rangle
\eeq 
where $\langle\bar{\Psi}\Gamma_{\chi}\Psi \rangle$ is computed using a path integral with weight $e^{-S_K}$. We can also define an axial current 
\beq
\mathcal{J}_{\mu}^{\chi}=\bar{\Psi}\Gamma_{\mu}\Gamma_{\chi}\Psi
\eqn{Jdef}
\eeq
 where $\Gamma_{\mu}=i\partial\widetilde K(p)/\partial{p^{\mu}}$, where we use the tilde to indicate the operator has been Fourier transformed to momentum space. Assuming that $\CD = \slashed{\partial}+\ldots$, where the ellipsis denotes nonderivative terms, then $\Gamma_\mu = \sigma_1\otimes\gamma_\mu$, in the notation where $\Gamma_\chi = \sigma_3\otimes 1$.  Applying the axial transformation $\Psi\rightarrow e^{i\Gamma_{\chi}\theta(x)}\Psi$, we can derive the Ward-Takahashi identity for this axial current as
\beq
\partial^{\mu}\mathcal{J}_{\mu}^{\chi}=2M\bar{\Psi}\Gamma_{\chi}\Psi
-\CA\ ,
\eqn{WI}\eeq
where the first term on the right is the classical divergence due to the mass $M$, and the second term is the anomaly, which can be computed using the method of Ref.~\cite{fujikawa1979path},
\beq
\CA=-2\,\lim_{\Lambda\rightarrow\infty} \Tr\left(\Gamma_{\chi}e^{K^2/\Lambda^2}\right) =-2 \CI(\infty)\ .
\eqn{anom}
\eeq
As we show below the anomaly vanishes in the cases we consider.  With $\CA=0$, \eq{WI} can then be used to express the index $\text{ind}(\CD)$ as
\beq
\text{ind}(\CD) = -\half \lim_{M\to 0} \int d^{d+1}x\, \partial_\mu \vev{ \CJ_{\mu}^{\chi}}\ .
\eqn{indform}
\eeq

The current $ \CJ_{\mu}^{\chi}$ is what we refer to as the generalized Hall current. With the anomaly $\CA$ vanishing, the nonzero divergence of this current in the $M\to 0$ limit arises from infrared divergences in the theory, and its inflow counts gapless edge states in the original Minkowski theory.

The generalized Hall current in \eq{indform} in the $M\to 0$ limit can be computed from a one-loop Feynman diagram $\Tr \Gamma_\mu \Gamma_\chi K^{-1}$, which may need to regulated.  The UV regulator will in general contribute to the current in odd spacetime dimensions, but not to its divergence, which arises from infrared physics.  However, in these cases the topological meaning of the current is obscured  if regulator contributions are neglected, as the Feynman diagram can be interpreted as being proportional to the winding number of a map from momentum space to an $n$-sphere -- but only if the momentum space of the fermion is compact. This makes the winding number sensitive to the ultraviolet behavior of the fermion propagator.  
Here we use Pauli-Villars regularization when required, computing the index of   $\CD_{\text{reg}}=\frac{\CD(m)}{\CD(\Lambda)}$, and then sending the  regulator mass  $\Lambda$   to infinity.  The regulated current then is given by  $\Tr \Gamma_\mu \Gamma_\chi K_\text{reg}^{-1} $, where $K_\text{reg}$ is given by \eq {kdef} with $\CD$ replaced by $\CD_{\text{reg}}$. We do this in any spacetime dimension $d+1$, but find that for even $d+1$, the regulator does not contribute to the index.

 In the next sections we work out in detail the examples of Ref.~\cite{kaplan2021index}, starting with the continuum example of a single Majorana fermion chain in $1+1$ dimensions with $0+1$ dimensional defect hosting a Majorana zero mode. This is followed by the example of multiple flavors of Majorana fermions in $1+1$ dimension with time reversal symmetry violation. In both cases, the generalized Hall current correctly produces the index of the fermion operator and counts the number of massless fermions in Minkowski space. We then discuss Dirac fermion in $2+1$ dimensions, which exhibits  the Integer Quantum Hall Effect, with the in-flow anomaly current described in Ref.~ \cite{Callan:1984sa}.  We then add a fermion number violating Majorana mass term this theory, modeling a topological superconductor in $2+1$ dimensions. Again we show how the divergence of the generalized Hall current correctly detects edge states, even though there is no conventional Hall current in such a system, and we discuss the topological origins of the index.  
After discussing the 2+1 dimensional case, we work through an analogous example in $3+1$ dimensions: that of Dirac fermion with a domain wall in its mass which describes three dimensional topological insulator and its edge states. We again compute the generalized Hall current, show how it reproduces the index and display it topological origins.  The final section provides a qualitative argument in the context of $1+1$ dimensional  interacting Majorana chains that the divergence of the generalized Hall current detects zeromodes even in the presence of interactions, even though the index of the free fermion propagator is no longer relevant.


\section{Majorana fermion $1+1$ dimensions} 
\label{sec:3}

\subsection{One flavor of Majorana fermion}

For our first example we consider a single Majorana fermion in $1+1$ dimension with domain wall profile for the Majorana mass, with the $0+1$ dimensional defect hosting a 1-component real massless fermion. Majorana edge states were first discussed in Refs. \cite{Kaplan:1999jn} and \cite{kitaev2001unpaired}. In this theory 
  there  is no continuous symmetry, and hence no conserved current in the bulk, and no chirality or chiral anomaly on the $0+1$ dimensional defect.    Therefore  this system cannot exhibit the conventional Hall response.  As we will show, however, we can compute generalized Hall currents which converge on the defect when it hosts gapless states.

The Minkowski Lagrangian in this case is 
\beq
\CL_M = \half\psi^T C \left(i\slashed{\partial}- m \right)\psi,
\label{lag1}
\eeq
which is the continuum version of the 1-flavor theory considered in Ref.~\cite{Fidkowski:2009dba}.
Here $\psi$ is a real, two-component Grassmann spinor.
The $\gamma$ matrices we  take are 
\beq
\gamma^0= C = \sigma_2\ ,\qquad 
\gamma^1 = -i\sigma_1\ ,\qquad
\gamma_\chi = \sigma_3\ , 
\eeq
where $\sigma_i$ are the Pauli matrices. This model of Majorana fermions has a rich phase structure when interactions are included \cite{Fidkowski:2009dba}, but in this section we restrict ourselves to analyzing the free theory. To construct the generalized Hall current for this model we first define the Euclidian theory\footnote{We use the mostly minus convention for the Minkowski metric, and in $d+1$ dimensions we denote both Minkowski and Euclidean time by $x^0$. The relation between Euclidian and Minkowski $\gamma$-matrices is $\gamma^0_M=\gamma^0_E$,  $\gamma^i_M=i\gamma^i_E$ and $\gamma_\chi$ is the same in both Minkowski and Euclidean spacetime.}
\beq
\CL_E =\half  \psi^T C \CD\psi\ ,
\eeq 
where in Euclidean spacetime
\beq
\CD = \slashed{\partial} + m\ ,\qquad \gamma_0 = C=\sigma_2\ ,\quad  \gamma_1=-\sigma_1\ , \quad \gamma_\chi=\sigma_3\ .
\eeq
We now wish to study the index of $\CD$, somewhat modified (there being a one-to-one correspondence between zeromodes of $\CD$ and $C\CD$).   The modifications involve replacing the step function mass  with a general scalar field $\phi_1(x)$, and adding a pseudoscalar field $\phi_2(x)$ as our diagnostic field, so that the more general operator we consider is
\begin{equation}
\begin{aligned}
\CD&=\slashed{\partial} + \phi_1+i\phi_2\gamma_{\chi} &= &\begin{pmatrix}\phi_1 + i \phi_2 & -i\partial_0-\partial_1 \\  i\partial_0 -\partial_1 &\phi_1 - i \phi_2\end{pmatrix} \ ,\\
\CD^\dagger &=-\slashed{\partial} + \phi_1-i\phi_2\gamma_{\chi} &= &\begin{pmatrix}\phi_1 - i \phi_2 & i\partial_0+\partial_1 \\  -i\partial_0 +\partial_1 &\phi_1 +i \phi_2\end{pmatrix} \ .
\end{aligned}
\end{equation}

To understand the logic of this construction, first consider the operator we are ultimately interested in, where $\phi_2=0$  and $\phi_1(x) = m_0\epsilon(x_1)$ with $m_0>0$.  In this case we find two solutions each to the equations $\CD\psi=0$ and $\CD^\dagger\chi=0$, namely
\begin{equation}
\begin{aligned}
\psi_-(x) &=  \begin{pmatrix}
1 \\ 1
\end{pmatrix}e^{m_0|x_1|}\ ,&\qquad
\psi_+(x) &=  \begin{pmatrix}
\phantom{-}1 \\ -1
\end{pmatrix}e^{-m_0|x_1|}\ ,\\
 \chi_-(x) &= \begin{pmatrix}
1 \\ 1
\end{pmatrix}e^{-m_0|x_1|}\ ,&\qquad
 \chi_+(x) &= \begin{pmatrix}
\phantom{-}1 \\ -1
\end{pmatrix}e^{m_0|x_1|}\ ,
\end{aligned}
\eqn{d2DWF}\end{equation}
 where the $\pm$ subscript refers to the eigenvalue of $\gamma_1$. None are normalizable since they are constant in $x_0$, whether or not they are localized in $x_1$ about $x_1=0$, and therefore the index calculation will give us  $\text{ind}(\CD)=0$. However, if we now add $\phi_2 = \mu_0 \epsilon(x_0)\ $  with $\mu_0>0$, a domain wall in Euclidian time,  we find the four solutions to $\CD\psi=0$ and $\CD^\dagger \chi=0$ are
 \begin{equation}
\begin{aligned}
\psi_-(x) &=  \begin{pmatrix}
1 \\ 1
\end{pmatrix}e^{m_0|x_1|+\mu_0|x_0|}\ ,
&\qquad
\psi_+(x) &=  \begin{pmatrix}
\phantom{-}1 \\ -1
\end{pmatrix} e^{-m_0|x_1|-\mu_0 |x_0|}\ ,\\
 \chi_-(x) &= \begin{pmatrix}
1 \\ 1
\end{pmatrix}e^{-m_0|x_1|+\mu_0|x_0|}\ ,
&\qquad
 \chi_+(x) &= \begin{pmatrix}
\phantom{-}1 \\ -1
\end{pmatrix}e^{m_0|x_1|-\mu_0|x_0|}\ ,
\ (x_2\to x_0)
\end{aligned}
\end{equation}
$\CD$ now has a single normalizable zeromode $\psi_+$ localized at $x_0=x_1=0$ while $\CD^\dagger$ has none.  Therefore the index is given by  $\text{ind}(\CD)=1$ in this background scalar field. By considering the other   signs for $m_0$ and $\mu_0$ one sees that more generally we get $\text{ind}(\CD)= -\nu_\phi$, where $\nu_\phi$ is the winding number of $\phi = \phi_1+i\phi_2$ in the $x_0-x_1$ plane.\footnote{This is easier to see if one replaces the singular step functions with something smoother, such as a $\tanh$ function. In fact, it has to be smoothed out to justify the derivative expansion we perform next.} As we shall see, this expectation is born out by the explicit calculation we give below.  Furthermore, our results persist even no matter how small we take $|\mu_0|$.  We also see how the zeromodes we are counting would not have existed if there hadn't been a solution to $\CD\psi=0$ for the case of interest, $\phi_2=0$ and $\phi_1 = m_0\epsilon(x_1)$ (for either sign of $m_0$), and so the nonzero value for  $\text{ind}(\CD)$  informs us that the original Minkowski theory has massless edge states.

Following the procedure laid out in the previous section, we compute the index of $\CD$ by constructing the fermion operator $K$ which serves as the fermion operator in a theory with twice the number of fermion degrees of freedom,
\beq
K =\begin{pmatrix}
0 && -\CD^{\dagger}\\
\CD && 0
\end{pmatrix} =\Gamma_\mu \partial_\mu + i (\phi_2 \Gamma_2 + \phi_1 \Gamma_3)\ .
 \eeq
where we have defined the five $4\times 4$ matrices
\begin{equation}
\begin{aligned}
\Gamma_i &= \sigma_1\otimes \gamma_i\ ,\  
&\Gamma_2 &= \sigma_1\otimes \gamma_\chi\ , \cr 
\Gamma_3 &= -\sigma_2\otimes 1\ , \  
&\Gamma_\chi &= \sigma_3\otimes 1\ .
\end{aligned}
\eqn{Gamdef2}
\end{equation}
with $i=0,1$.

Our task is to compute the part of the chiral current $\CJ_\mu = \bar\Psi \Gamma_\mu \Gamma_\chi \Psi$ that contributes to the index, where $\mu=0,1$ and $\Psi$ is a fermion with action $S = \bar\Psi K \Psi$.  To do this we first write the scalars as
\beq
\phi &=& \phi_1 + i \phi_2 = (v+\rho(x)) e^{i\theta(x)}
\eqn{scalar}
\eeq
assuming  constant $v$ with $\rho$ and $\theta$   slowly varying about $ \rho=\theta = 0$, their gradients falling off at infinity.    To compute the part of the current that contributes to the index we need the leading term in a $1/v$ expansion, since higher order terms will drop off too fast at infinity to contribute to the integral $\int \partial_\mu \CJ_\mu$.  Thus we can write
\beq
K &=& K_0 + \delta K\ ,
\eeq
with
\beq
 K_0 = 
 \partial_\mu \Gamma_\mu + i v\, \Gamma_3\ ,\qquad
\delta K =
i v\, \theta(x)  \Gamma_2 + i   \rho(x) \Gamma_3\
\ .
\eqn{Kdef2}\eeq
Then $K_0^{-1}$ will be our free fermion propagator, while we perturb in $\delta K$.

\begin{figure*}[t]
\centerline{\includegraphics[height=3 cm]{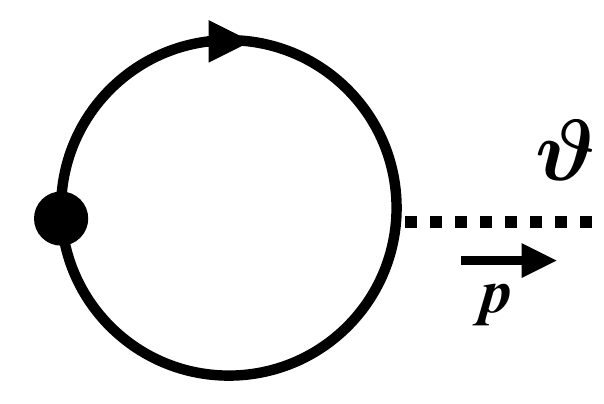}}
\caption{\it The loop diagram for computing the generalized Hall current for the   $1+1$-dimension Dirac fermion. The black dot is an insertion of the chiral current $\Gamma_\mu\Gamma_\chi$ with incoming momentum $p_\nu$. The outgoing field $\alpha$ is the spatially varying part of the phase of the complex field $\phi_1+i\phi_2$, and the fermion propagator is given by $K_0^{-1}$.}
\label{fig:2dloop}
\end{figure*}

When expanding $\CJ_\mu$  in $\delta K$, note that because of the insertion of $\Gamma_\chi$ in the fermion loop we require that the rest of the graph supplies one each of the other four $\Gamma$ matrices in order to get a nonzero contribution from the trace.   The matrices $\Gamma_{0,1,3}$ can be supplied by the fermion propagators $1/K_0$, while $\Gamma_2$  must arise from an insertion of $\theta$.    Thus the leading contribution is given by the graph in  Fig.~\ref{fig:2dloop} expanded to linear order in the momentum carried  by $\theta$.  

Using $\tilde K_0$ to denote the Fourier transform of $K_0$ in momentum space, the diagram can be computed as  
 \beq
\CJ_\mu&=&v\, \frac{\partial\theta}{\partial x_\nu}\int \frac{d^2q}{(2\pi)^2}\,\Tr\Biggl[
\Gamma_\mu\Gamma_\chi 
 \left(\frac{\partial  \tilde K_0^{-1}}{\partial q_\nu} \right)\Gamma_2\tilde K_0^{-1} \Biggl]\cr
 &=&
 \epsilon_{\mu\nu} \partial_\nu \theta \int \frac{d^2q}{(2\pi)^2}\frac{4v^2}{(q^2+v^2)^2} \cr 
 &=&
  \frac{1}{\pi} \epsilon_{\mu\nu} \partial_\nu \theta  \ .
\eqn{J11}
\eeq
In this expression,  the derivative $\partial_\nu$ acting on $\theta(x)$ is with respect to $x_\nu$, while the derivative $\partial_\nu$ acting on $\tilde K_0$ is with respect to $q_\nu$.  The positive sign in the first line above arises from $(-1)$ from the fermion loop, a $(-i)$ from Fourier transforming $p_\nu\to -i \partial/(\partial x_\nu)$, and another $(-i)$ from the   $\theta$ vertex factor $-i v \Gamma_2 $. 

From \eq{indform} it follows that the index of $\CD$ is given by
\beq
\text{ind}(\CD) = -\half\int d^2x \partial_\mu \CJ_\mu  =-\frac{1}{2\pi} \oint \frac{\partial\theta}{\partial x_\mu} d\ell_\mu = -\nu_\phi\ ,
\eqn{ind2a}\eeq
where $\nu_\phi$ is the winding number of $\phi$. This result agrees with what was predicted from our earlier heuristic argument with a configuration of crossed domain walls.

Before turning to the question of topology we address two issues about the way we handled the Ward-Takahashi identity for the generalized Hall current, given in \eq{WI}.  First of all, we set $M=0$ from the outset, rather than performing the computation at nonzero $M$ and then taking the limit $M\to 0$ at the end.  This is discussed further in \S~\ref{sec:interactions}, but in brief: the role of $M$ was to serve as an IR regulator for the calculation, but instead we used our background field $\phi$ to serve as the IR regulator.  Note that the generalized Hall current we found in \eq{J11} is proportional to $\partial_\mu\theta = i\phi^* \overleftrightarrow{\partial_\mu}\phi/2|\phi|^2$, and that the inverse dependence on $|\phi|$ indicates its role as an IR regulator.  Secondly, we ignored the anomaly term, $\CA$.  This quantity can be computed using the methods of Fujikawa \cite{fujikawa1979path}, where
\beq
\CA = \lim_{\Lambda\to\infty} \Tr \Gamma_\chi e^{K^2/\Lambda^2}\ .
\eeq
(Note that $K$, defined in \eq{Kdef2},  is antihermitean.) The two-derivative term in $K^2$ gives rise to the Gaussian integral
\beq
\int \frac{d^2k}{(2\pi)^2} \, e^{-k^2/\Lambda^2} \propto \Lambda^2\ .
\eeq
On the other hand, since the $\Gamma$-matrices obey the Clifford algebra for $SO(4)$, the $\Gamma$-matrix trace with $\Gamma_\chi$ requires that the expansion of $e^{K^2/\Lambda^2}$ supplies at least four different $\Gamma_\mu$ to the trace before one obtains a nonzero value.  As $K/\Lambda$ is linear in the $\Gamma$ matrices, these four $\Gamma$ matrices will be accompanied by a factor of $1/\Lambda^4$ which overwhelms the factor of $\Lambda^2$ from the momentum integral and causes the anomaly term $\CA$ to vanish as $\Lambda\to \infty$.  This will occur in all of our examples, since the doubled theory always has the same  $d+1$  spacetime dimension as the original theory, while the $\Gamma$-matrices belong to the Clifford algebra for $SO(d+3)$.

 The calculation we have performed does not make explicit the topological quantization of the index, since it seems like an accident of the momentum integral that the integral the divergence of the generalized Hall current came out to be an integer times $\nu_\phi$.  To display  the topology underpinning of the calculation explicitly we  define  $\tilde \CD_0$ to equal $\tilde \CD$ with  $\phi_1 \to v$ and $\phi_2\to 0$ and use the identities $\Gamma_\mu = -i\partial_\mu \tilde K_0$ and   $\{\gamma_\chi,\tilde\CD_0\} = 2 v\gamma_\chi$ to rewrite \eq{J11} as 
\beq
\CJ_\mu&=&-iv\partial_\nu\theta\, \int \frac{d^2q}{(2\pi)^2}\,\Tr\Biggl[
\Gamma_\chi 
 \left(\partial_\nu  \tilde K_0^{-1} \right)\Gamma_2 \tilde K_0^{-1} \partial_\mu \tilde K_0\Biggr]
 \cr
&&
\cr
&=&
-i v\partial_\nu \theta \, \int \frac{d^2q}{(2\pi)^2}\,\Tr\Biggl[\gamma_\chi\left( \tilde D_0^{-1} \partial_\mu \tilde D_0\partial_\nu \tilde D^{-1}_0 
+\tilde D_0 \to \tilde D^\dagger_0\right)\Biggr]
\cr &&\cr
&=&
\frac{i}{4} \epsilon_{\mu\nu} \partial_\nu \theta \epsilon_{\sigma\tau} \, \int \frac{d^2q}{(2\pi)^2}\,
\Tr\Biggl[\gamma_\chi\left( \tilde D_0^{-1} \partial_\sigma \tilde D_0\tilde D_0^{-1} \partial_\tau \tilde D_0 + 
 \tilde D_0\partial_\sigma \tilde D^{-1} _0\tilde D_0\partial_\tau \tilde D^{-1} _0 +\tilde D_0 \to \tilde D^\dagger_0\right)\Biggr]\cr &&
\eqn{D3}\eeq
 We can then define the special unitary matrix
\beq
\xi= \frac{\tilde \CD_0 }{\sqrt{\det\tilde \CD_0}} = \frac{v }{\sqrt{q^2+v^2}} +i \hat q_\mu\gamma_\mu\frac{q}{\sqrt{q^2+v^2}} \ ,
\eeq
in terms of which \eq{D3} can be rewritten as
\beq
\CJ_\mu=\frac{i}{2}\epsilon_{\mu\nu}\partial_\nu \theta\,\epsilon_{\sigma\tau}\int \frac{d^2q}{(2\pi)^2}\Tr\gamma_\chi\left[\left(\xi^\dagger \partial_\sigma \xi\right)\left( \xi^\dagger\partial_\tau  \xi \right)+\left(\xi \partial_\sigma \xi^\dagger\right)\left(\xi \partial_\tau  \xi^\dagger\right)   \right]\ ,
\eqn{D3new}\eeq
unaffected by the normalizing determinant factor (which can't vanish for nonzero $v$).  As a final manipulation, we can define the ``axial current'',
\beq
A_j = \frac{i}{2}\left(\xi^\dagger \partial_j \xi - \xi \partial_j \xi^\dagger\right)\ ,
\eeq
and using the identity $\gamma_\chi \xi \gamma_\chi = \xi^\dagger$ we can write the generalized Hall current as
\beq
\CJ_\mu=-i\epsilon_{\mu\nu}\partial_\nu \theta\,\epsilon_{\sigma\tau}\int \frac{d^2q}{(2\pi)^2}\Tr\left[ \gamma_\chi A_\sigma A_\tau   \right]\ .
\eqn{D3newA}\eeq
In  Appendix~\ref{sec:appendix_topology} we show that the above integral is proportional to the winding number of the map provided by the Dirac propagator from momentum space compactified to $S^2$, to ``Dirac space'' (the analog of the Bloch sphere), also $S^2$ in two spacetime dimensions. That winding number $\nu_q$ is an element of $\pi_2(S^2) = {\mathbb Z}$, and for this particular map we have $\nu_q=1$.  Making use of the normalization given in \eq{rtg} we arrive at

\beq
\CJ_\mu = \frac{\epsilon_{\mu\nu}\partial_\nu\theta}{\pi} \,\nu_q\ ,\qquad \nu_q=1\ .
\eeq

The topological quantization in momentum space  is analogous to the topological origin of the quantization of the Integer Quantum Hall Effect discovered in the celebrated TKNN paper and related work \cite{thouless1982quantized,niu1984quantised,niu1985quantized}.

Our final result for the index is then given by
\beq
\text{ind}(\CD) =   -\nu_\phi  \nu_q\ ,
\eeq
where $\nu_\phi$ is the winding number of our diagnostic field in coordinate space, obtained by integrating $\epsilon_{\mu\nu}\partial_\nu\theta$ and   determined by the homotopy group $\pi_1[U(1)]= \mathbb Z$. We see that the index, which counts edge states, is manifestly quantized by simultaneous nontrivial  topology in both momentum and coordinate space.
This result confirms our heuristic argument about localizing the zeromode with crossed domain walls, and agrees with \eq{ind2a}.

Relativistic quantum field theories can be assigned to the same ten topological classes as used for condensed matter systems; for example see Ref.~\cite{PhysRevB.78.195125}.  The translation is simple, with what are called time reversal symmetry, particle hole symmetry, and sublattice symmetry replaced by the conventional $T$, $C$ and $CT$ symmetries respectively in the relativistic theory.  The ten categories are constructed from (i)   time reversal symmetry  with $T^T=T$, or $T$ symmetry with $T^T=-T$, or $T$ violation; (ii)  charge conjugation symmetry  with $C^T=C$, or $C$ symmetry with $C^T=-C$, or $C$ violation; (iii) both $C$ and $T$ violation, but good $CT$ symmetry.   The only caveat in comparing  in using the tables from condensed matter papers is that the symmetry of $C$ is opposite that of the conventional particle-hole symmetry ``$P$", since in the relativistic case charge conjugation is the transformation $\psi \to C\bar\psi^T$ for a Dirac fermion, while particle hole symmetry is $\psi\to P \psi^* $, without the extra $\gamma^0$ matrix.  In the present example we have both $T$ and $C$ with symmetric $T$ and antisymmetric $C$, which puts the system in the BDI class, with topological invariant ${\mathbb Z}$. If we had considered a $N_f$ -flavor version of the theory considered in this section without special global flavor symmetries, we would have trivially found $   \text{ind}(\CD) =   -N_f\nu_\phi  \nu_q$, which can take on any value in ${\mathbb Z}$ as one would expect for a BDI topological class in one spatial dimension.


\subsection{Multiple flavors of $d=1+1$ Majorana fermions with time reversal symmetry violation}
 \label{sec:3Nf}
 
With $N_f$ flavors of free fermions in our $1+1$ dimensional model, it is possible to include both scalar and pseudoscalar mass terms:
\beq
\CL_M = \half\psi_i^T C \left(i\slashed{\partial}\delta_{ij}- m_{ij} - i\gamma_{\chi} \mu_{ij} \right)\psi_j,
\eqn{lagNf}
\eeq
where $m$ must be a real and symmetric matrix, while $\mu$ is imaginary and antisymmetric. Without loss of generality it is possible  to take $m_{ij}$ to be diagonal. 

The theory with $\mu_{ij}=0$ is invariant under the antiunitary time-reversal symmetry, $\psi_i(x,t) \to \sigma_1\psi_i(x,-t)$.  When the  masses have domain wall profiles, such as  $m_i = m_0\epsilon(x)$ with $m_0>0$, there will be $N_f$ massless Majorana modes localized at the mass defect with wave function 
\beq
\eta_i(t) \begin{pmatrix} \phantom{-}1\\ -1\end{pmatrix}\qquad 
\eeq
in the $0+1$ dimension theory at the defect.
 In order to gap these modes in the free theory, one might add a term $\mu_{ij} \eta_i \sigma_1 \eta_j$ with $\mu$ imaginary and antisymmetric, and it is easy to see that this term lifts into the bulk theory as the $\mu$ term in \eq{lagNf}.
However, the pseudoscalar $\mu$ term is odd under the time-reversal symmetry we identified (since $\mu$ is imaginary), and therefore time reversal symmetry requires $\mu=0$ and ensures that the edge states remain gapless.    Such a system is in the BDI topological class, whose gapless states are characterized by the group $\mathbb Z$ \cite{wiki:periodic}.

For $N_f \ge 3$ it is possible to show that one can choose $m$ and $\mu$ such that time reversal symmetry is broken, and generically the edge states will be gapped pairwise, so that there will be one massless edge state for $N_f$ odd, and none for $N_f$ even.  Such a system is in the D class, with topology characterized by the group ${\mathbb Z}_2$ \cite{wiki:periodic}.

   For $N_f=2$, time reversal is  not actually broken when $\mu\ne 0$, since the theory is invariant under the simultaneous antiunitary transformations $\psi_1(x,t) \to+ \sigma_1\psi_1(x,-t)$ and $\psi_2(x,t) \to- \sigma_1\psi_2(x,-t)$. In this case edge states can still be gapped because the fermions transform with opposite signs under time reversal and the system is topologically trivial. So the system is indistinguishable from the D class as far as edge states are concerned.
   
 We focus on this simplest $N_f=2$ case to show how the index calculation procedure we have developed here gets the correct answer, that there are no gapless edge states.  The Minkowski theory we consider is
\beq
\CL_M = \half\psi_i^T C \left(i\slashed{\partial}- m_0\epsilon(x) - i\gamma_{\chi} \mu\tau_2 \right)\psi\ ,
\label{lagNf2}
\eeq
 where we have suppressed the two flavor indices. The $(i\slashed{\partial}- m_0\epsilon(x))$ operator is diagonal in flavor, while $\tau_2$ is the $y$-Pauli matrix acting in flavor space, and $\mu$ is real and constant in spacetime.
 
 Our Euclidean operator $\CD$ with a diagnostic field $\phi(x)$ in this case is given by
 \beq
 \CD = \slashed{\partial} + \phi_1(x) + i \left(\phi_2(x) + \mu \tau_2\right)\gamma_\chi\ .
 \eeq
 This can be diagonalized in flavor to give two 1-flavor Dirac operators
 \beq
 \CD_\pm =  \slashed{\partial} + \phi_1(x) + i\left( \phi_2(x) \pm \mu\right)\gamma_\chi\ .
 \eeq
 Now we expect the sum of two contributions to $\CJ_\mu$ proportional to $\epsilon_{\mu\nu} \partial _\nu\left(\theta_+ + \theta_-\right)$ where
 \beq
 \theta_\pm = \arctan \frac{\phi_2(x) \pm \mu}{\phi_1(x)}\ .
 \eeq
 Now we see that even if $\phi(x) = \phi_1(x) + i \phi_2(x)$ has winding number, as we take the limit $\phi_2\to 0$, that winding number vanishes.  For example, if we take $\phi_1 = m \epsilon(x)$ and $\phi_2(x) = m' \epsilon(\tau)$, the two contributions to the index will both equal one for $|m'|>|\mu|$, but will jump to zero for $|m'|<|\mu|$ as we remove our diagnostic field -- this is possible because the bulk goes gapless at the critical value $|m'|=|\mu|$.  So the index in this case would give the correct answer of zero, as it will for any even number of flavors.
 
 If we have an odd number of flavors with a constant $\mu$ matrix for $N_f= 2n+1$, we will find $n$ pairs of fermions coupling to $\phi_2(x)$ with $\pm\mu_i$ shifts, each contributing zero to the index; however there would be one flavor with no shift, and we would find an index of one then.  Thus we find that $\text{ind}(\CD)$ takes values in ${\mathbb Z}_2$. That is the correct answer since having no $T$ symmetry    while having $C$ symmetry with antisymmetric $C$ puts the model in the $D$ topological class, whose topological invariant in one spatial dimension is ${\mathbb Z}_2$.

 We emphasize though that in our calculation the reduction of the topological invariant from ${\mathbb Z}$ in the time reversal symmetric case to ${\mathbb Z}_2$ in the case with broken time reversal symmetry manifests itself in the change in the spacetime coordinate topology that the fermion sees, and not due to a change in momentum space.  The index we compute sees the topology in phase space, and is sensitive to both.


 \section{Dirac fermion and Majorana fermions in 2+1 dimensions}
\label{sec3}
  

  \subsection{A $d=2+1$ Dirac fermion with $U(1)$ fermion  number symmetry}
  
Our next example is a massive Dirac fermion in 2+1 (Minkowski) dimensions, which  is directly analogous to the Integer Quantum Hall Effect and is quite familiar.  The coordinates are  $\{x^0, x^1, x^2\}=\{t,x,y\}$  and the fermion mass  $m(y)$ which has a ``domain wall'' structure -- a monotonically increasing function of $y$ with $m(0)=0$   \cite{Jackiw:1975fn}.   
For convenience we choose the particular basis for the $\gamma$ matrices
\beq
\gamma^0 = \sigma_2\  ,\quad \gamma^1 = -i\sigma_1\ ,\quad \gamma^2 = i\sigma_3\ ,
\eeq
in which case the Dirac equation $\left[i\slashed{\partial} - m(y)\right]\psi = 0$ has the two special  solutions,
\beq
\psi_\pm = e^{-i \omega (t \pm x)} e^{\mp \int^{y}_0 ds \, m(s)}\, \chi_\pm \ ,
\eeq
where $\chi_\pm$ are constant 2-component spinors satisfying $\sigma_3 \chi_\pm = \pm \chi_\pm$.
The solution $\psi_+$ is localized on the 1+1 dimensional domain wall and corresponds to a massless Weyl fermion that travels at the speed of light in the $-x$ direction.   However, since  $ e^{+ \int^{y}_0 ds \, m(s)}$ is not normalizable, the $\psi_-$ solution does not correspond to a state in the Hilbert space. Therefore the spectrum on the domain wall is chiral, and if fermion number is gauged, this $1+1$ dimensional theory on the domain wall is anomalous. 

All other eigenstates of the Dirac operator are gapped and are not localized. One might therefore expect that these heavy states could be integrated out, leaving an effective $1+1$ dimensional theory of a Weyl fermion at $y=0$, along with irrelevant operators.  However, when fermion number symmetry is gauged the heavy fermions do not decouple entirely, giving rise to a marginal Chern-Simons operator after being integrated out of the theory, as pointed out by a number of authors \cite{redlich1984gauge,ishikawa1984chiral,Callan:1984sa}. For a constant mass $m$, this contribution is \footnote{Our notation is that the covariant derivative is $D_\mu =( \partial_\mu - i A_\mu)$, where $A_\mu$ has mass dimension $1$, setting the electric charge to $e=1$.}
\beq
\CL_\text{CS} = \frac{1}{4\pi} \frac{m}{|m|}\epsilon^{\mu\nu\rho} A_\mu\partial_\nu A_\rho\ .
\eqn{Lcs}
\eeq
The coefficient of this operator depends on the sign of  $m$ but  not its magnitude; the dependence on the sign is required because $m$ changes sign under $T$ or $P$ transformations (time reversal and space reflection), as does the Chern-Simons operator.  Variation of $\CL_{CS} $ with respect to the gauge field gives rise to a current
\beq
J^\mu_\text{CS} = \frac{1}{2\pi} \frac{m}{|m|}\epsilon^{\mu\nu\rho}  \partial_\nu A_\rho 
\eqn{Jcs}
\eeq

Callan and Harvey \cite{Callan:1984sa}  discussed the effects of this operator in the presence of a domain wall profile, substituting  the step function $ \epsilon(y)$ for $m/|m| $ in the above expression, so that $J^\mu_\text{CS}$ corresponds to a current flowing on or off of the domain wall from both sides, with a divergence proportional to $\delta(y)$ that exactly cancels the anomalous divergence of the chiral current on the domain wall.
 There are a few problems with their analysis.  For one, the substitution of spatially varying mass  for a constant mass in the coefficient of the Chern Simons term is not correct where the fermion mass is varying appreciably, i.e. near the domain wall;  however to demonstrate overall charge conservation, one need only look at the spatial integral of the divergence of the current, which involves the Chern Simons coefficient at infinity, where the substitution is valid if the mass is not rapidly changing there. A second  problem is that while the 1-loop integral for deriving $\CL_{CS}$ is finite, the full theory requires a regulator which contributes to the Chern-Simons operator as well.  For example, with a Pauli-Villars regulator of mass $\Lambda$, the coefficient in \eq{Lcs} and \eq{Jcs} becomes proportional to $(m/|m| - \Lambda/|\Lambda|)$.  With the Pauli-Villars mass $\Lambda$ being independent of $y$, the effect is to double the inflowing current on one side of the domain wall, and cancel it on the other. This has no effect on the divergence of the current, but corrects the coefficient of the Chern-Simons term and makes a topological interpretation possible by compactifying momentum space. Other regularization schemes, such as the lattice, can have richer topological phase structure.

Even with the correct expression for the current, the connection with the Integer Quantum Hall Effect  is obscure, beyond the fact that an electric field in the $x$ direction gives rise to a current in the $y$ direction; missing is an analog of the famous plot of the resistivity $\rho_{xy}$  versus magnetic field in the condensed matter system, with its characteristic steps. In the Dirac fermion case, the domain wall plays the role of one of the boundaries of the Quantum Hall system, and the Dirac mass --- which is time reversal violating in $2+1$ dimensions --- plays the role of the magnetic field, while the resistivity is the ratio of the applied electric field to the Chern Simons current.  However, one does not see a stepwise increase in the current as a function of the Dirac mass.   Once again, that is a result of how the short-distance physics of the theory is being regulated.  For example, when one uses a lattice and Wilson terms to regulate the UV, one does see quantized  jumps in the Chern-Simons current as a function of the fermion mass \cite{Kaplan:1992bt,Jansen:1992tw,Golterman:1992ub}.

\begin{figure*}[t]
\centerline{\includegraphics[height=3 cm]{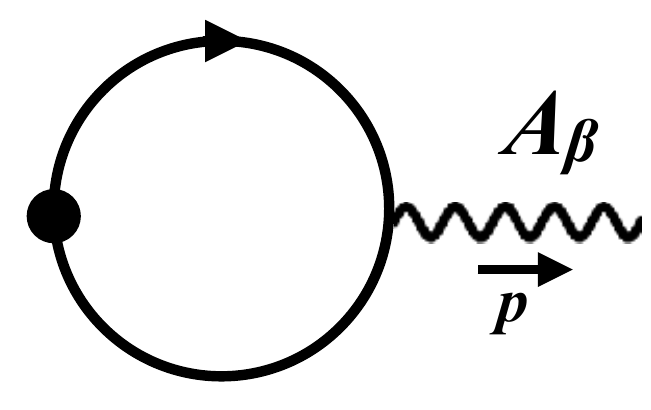}}
\caption{\it Loop diagram for computing the generalized Hall current for the   $2+1$-dimension Dirac fermion. The black dot is an insertion of the chiral current $\Gamma_\alpha\Gamma_\chi$ with incoming momentum $p$, and the propagators are given by $K^{-1}$.}
\label{fig:3dloop}
\end{figure*}

The  Euclidean fermion operator for this fermion is 
\beq
 \CD = \left[\slashed{D} + m(y)\right]\ ,\qquad \gamma_0 = \sigma_2\ ,\quad \gamma_1 = -\sigma_1\ ,\quad \gamma_2 = \sigma_3\ ,
 \eqn{d3basis}
 \eeq
with $D_\mu = \partial_\mu +i A_\mu$. From \eq{kdef} we have
\beq
K=\begin{pmatrix}
0 && -\CD^{\dagger}\\
\CD && 0
\end{pmatrix} = D_\mu \Gamma_\mu - i m(y) \Gamma_3\ ,
\eeq
where we define the matrices (satisfying the $SO(5)$ Clifford algebra)
\beq
\Gamma_\mu = \sigma_1\otimes\gamma_\mu\ ,\qquad \Gamma_3 = \sigma_2\otimes 1\ ,\qquad  \Gamma_\chi = \sigma_3\otimes 1
\eeq
for $\mu=0,1,2$.
Our task now is to compute the generalized Hall current $\CJ_\mu$  in \eq{Jdef}.  Since the bulk mass $|m|$ serves to regulate the IR divergences of the theory, we can take the limit $M\to 0$ in \eq{indform} from the start. Another simplification is that since we only need the current in-flow from infinity in the limit that the gauge field strength is weak, we can compute $\CJ_\mu$ to leading order in a $1/m$ expansion which is given by the Feynman diagram in Fig.~\ref{fig:3dloop} expanded to first order in the incoming momentum\footnote{The term we keep will give a finite contribution to the index, while higher order operators  fall off too fast at infinity to contribute to the integral $\int \partial_\mu \CJ_\mu$.}.    

We first perform the calculation naively, without a Pauli-Villars regulator,  with the result
\beq
\CJ_\alpha =  \partial_\gamma A_\beta \int \frac{d^3q}{(2\pi)^3} \, \Tr \Gamma_\alpha \Gamma_\chi (\partial_\gamma \tilde K_0^{-1}(q)) \Gamma_\beta  \tilde K_0^{-1}(q)\ ,
\eeq
where  as before, the tilde indicates a Fourier transform to momentum space, and $ \tilde K_0= \tilde K\vert_{A_\mu=0}$. The factor of $ \Gamma_\alpha \Gamma_\chi $ in the trace comes from the insertion of the generalized Hall current, while the photon vertex gives  $i\Gamma_\beta$; the two factors of $ \tilde K_0^{-1}$ are the two fermion propagators, and the derivative $\partial_\gamma=\partial/\partial q_\gamma$ arises from expanding the graph to first order in the external momentum $p$.   With $\Gamma_\alpha = -i \partial_\alpha \tilde K_0$, we can rewrite this as
\beq
\CJ_\alpha(p) &=&   \partial_\gamma A_\beta \int \frac{d^3q}{(2\pi)^3} \, \Tr \left[\Gamma_\chi \left(\tilde K_0^{-1} \partial_\gamma \tilde K_0\right) \left(\tilde K_0^{-1} \partial_\beta \tilde K_0\right) \left(\tilde K_0^{-1} \partial_\alpha \tilde K_0\right)\right]\cr
&=&
 -\epsilon_{\alpha \beta\gamma}\partial_\gamma   A_\beta  \int \frac{d^3q}{(2\pi)^3} \,\frac{4m}{(m^2+q^2)^2}\cr
& =& 
-\frac{1}{2\pi} \frac{m}{|m|}\epsilon_{\alpha \beta\gamma}\partial_\gamma   A_\beta   \ .\cr &&
\eqn{Kloop3}
\eeq
 With a domain wall profile for $m(y)$, the above expression exhibits generalized Hall current flow converging on (or diverging from) the wall similar to the electromagnetic current found by Callan and Harvey \cite{Callan:1984sa}.  

To better understand the topology behind the loop integral, it is convenient to rewrite \eq{Kloop3}  in terms of the fermion operator $\tilde \CD_0$ of the undoubled theory, 
\beq
 \CJ_\alpha 
&=&
 -\epsilon_{\alpha \beta\gamma}\partial_\gamma   A_\beta \left(\frac{1}{3}\epsilon_{ijk}\int \frac{d^3q}{(2\pi)^3}  \Tr \left[ \left(\tilde \CD_0^{-1} \partial_i\tilde\CD_0\right) \left(\tilde \CD_0^{-1} \partial_j \tilde \CD_0\right) \left(\tilde \CD_0^{-1} \partial_k \tilde \CD_0\right)\right]\right)\cr
 &&\cr &&
 \eqn{Dloop}\eeq

 As in the example in the previous section, for constant $m$ we can define the $SU(2)$ matrix
\beq
 U(\bfq) = \frac{\tilde \CD_0(\bfq)}{\sqrt{\det  \tilde\CD_0(\bfq)}}  \equiv \cos\frac{\theta}{2}+ i \hat {\boldsymbol \theta}\cdot{\boldsymbol\gamma}\sin\frac{\theta}{2}
\eqn{udef1}
\eeq
where ${\boldsymbol \theta}$ is a real 3-vector with $\theta= |{\boldsymbol \theta}| $ and
\beq
 \cos\frac{\theta}{2}= \frac{m}{\sqrt{m^2+q^2}} \ ,\qquad
\sin\frac{\theta}{2}=\frac{ q}{\sqrt{m^2+q^2}}\ ,\qquad  \hat {\boldsymbol \theta}=\hat{\bf q}\ . 
\eqn{udef2}\eeq
The coordinates $\theta_i$ parametrize $SU(2)\cong S^3$ as a three-dimensional ball of radius $2\pi$, where we can identify $ U=1$ and $ |{\boldsymbol \theta}|=0$ as the ``north pole'' of $S^3$, while $ U=-1$ and $ |{\boldsymbol \theta}|=2\pi$ corresponds to the ``south pole''. However, as the magnitude of the momentum $q$ ranges from $q=0$ to $q=\infty$, we see that $ U(\bfq)$ only visits half of $S^3$. In particular, for $m>0$ we see that  $q=0$ corresponds to  $|{\boldsymbol \theta}|=0$,  the north pole, while $q\to \infty$ corresponds to $  |{\boldsymbol \theta}| = \pi$, the equator, so the mapping only covers the northern hemisphere of $S^3$.  On the other hand, when $m<0$ one can see that the mapping just covers the southern hemisphere. We can rewrite our expression \eq{Dloop} in terms of the $\theta_i$ variables as
\beq
 \CJ_\alpha &=&-\frac{1}{3}  \epsilon_{\alpha \beta\gamma}\partial_\gamma  A_\beta   \,\epsilon_{ijk}\int \frac{d^3q}{(2\pi)^3} \, \Tr \left[\left( U^\dagger \frac{\partial  U}{\partial q_i }\right)
 \left(  U^\dagger \frac{ \partial  U}{\partial\ q_j}\right) 
 \left( U^\dagger\frac{\partial  U}{\partial q_k}\right)\right]\cr &&\cr 
&=&
-\frac{1 }{\pi} \epsilon_{\alpha \beta\gamma}\partial_\gamma  A_\beta \left(\frac{1}{24\pi^2}\epsilon_{ijk}\int_{V_{1/2} } d^3\theta\, \Tr \left[ \left( U^\dagger \frac{\partial  U}{\partial \theta_i }\right)
 \left(  U^\dagger \frac{ \partial  U}{\partial\theta_j}\right) 
 \left( U^\dagger\frac{\partial  U}{\partial\theta_k}\right)\right]\right)\ , 
\eqn{U3 }
 \eeq
 where as discussed above, the volume of integration $V_{1/2}$  only covers half of the ball, corresponding to the northern hemisphere of $S^3$ for $m>0$, or  the southern hemisphere for $m<0$. As such, the integral cannot be considered to be the winding number of a map from momentum space to $S^3$.  
  The situation is remedied when a regulator is included.  Here we consider a Pauli-Villars regulator, which corresponds to replacing $\CD \to \CD_\text{reg}=  \CD/\CD_\text{PV}=\CD(m)/\CD(\Lambda)$ everywhere, where in $\CD_\text{PV}$ we have simply replaced $m$ by the Pauli-Villars mass $\Lambda$, which we will take to $+\infty$ after the calculation. This has the effect of compactifying momentum space, since $\lim_{q\to\infty} \tilde \CD_\text{reg}(\bfq) = 1$, independent of the orientation of the momentum vector $\bfq$.
  The calculation proceeds as before, but now the $SU(2)$ matrix in \eq{udef1} gets replaced by
 \beq
 U_\text{reg}(\bfq) &=& \frac{ \tilde  \CD_\text{reg}(\bfq)}{\sqrt{\det  \tilde  \CD_\text{reg}(\bfq) }}    
 \equiv \cos\frac{\theta_\text{reg}}{2}+ i \hat {\boldsymbol \theta}_\text{reg}\cdot{\boldsymbol\gamma}\sin\frac{\theta_\text{reg}}{2}   \eeq
   where $ \hat {\boldsymbol \theta}_\text{reg}=\hat{\bf q}$ as before, and
   \beq
   \cos\frac{\theta_\text{reg}}{2}= \frac{\Lambda  m+q^2}{\sqrt{\left(m^2+q^2\right)
   \left(\Lambda ^2+q^2\right)}}\ ,
   \qquad
\sin\frac{\theta_\text{reg}}{2} =\frac{q (\Lambda -m)}{\sqrt{\left(m^2+q^2\right)
   \left(\Lambda ^2+q^2\right)}}\ .&&\cr &&
   \eeq
We note that at $q\to\infty$ we have $U_\text{reg}(\bfq)\to 1$, corresponding to compactifying momentum space to $S^3$ and mapping the point at $q=\infty$ to the north pole of the $S^3$ parametrized by ${\mathbb \theta}$. (This is in contrast to $ U$ in \eq{udef1} which maps the 2-sphere  at $q\to\infty$ onto the equator of $S^3$).  At $q=0$, however, we find that $U_\text{reg}(0)= \pm 1$ depending on the relative sign of $m$ and $\Lambda$. When $\Lambda$ and $m$ have the same sign, $\cos\frac{\theta_\text{reg}}{2}\ge 0$ for all values of $q$, with $\theta_\text{reg}=0$ for both $q=0$ and $q=\infty$.  In this case  $U_\text{reg}(q)$ describes a topologically trivial map from our compact momentum space to the northern hemisphere of $S^3$ ($0\le \theta_\text{reg}\le \pi/2$).  However, when $m$ and $\Lambda$ have opposite signs  $U_\text{reg}(q)$ is a nontrivial map from momentum space to $S^3$ with winding number equal to one.   It is no surprise then that we find that for the regulated theory 
\beq
 \CJ_\alpha &=&-\frac{1}{\pi} \epsilon_{\alpha \beta\gamma}\partial_\gamma  A_\beta(p)\left(\frac{1}{24\pi^2}\epsilon_{ijk}\int_{V}  d^3\theta_\text{reg}\,
\Tr \left[ 
\left( U_\text{reg}^\dagger \frac{\partial U_\text{reg}}{\partial \theta_i }\right)
 \left( U_\text{reg}^\dagger \frac{ \partial U_\text{reg}}{\partial\theta_j}\right) 
 \left(U_\text{reg}^\dagger\frac{\partial U_\text{reg}}{\partial\theta_k}\right)
 \right]
\right)\cr 
&=&
\frac{ \nu_q}{\pi} \epsilon_{\alpha \beta\gamma}\partial_\beta  A_\gamma \ , 
\eqn{Jreg}
 \eeq
 where
 \beq
 \nu_q = \half \left(\frac{\Lambda}{|\Lambda|} -\frac{m}{|m|}\right)  
 \eeq
is the winding number of the map from compact momentum space to $S^3$, where $\nu_q=0$ when $\Lambda$ and $m$ have the same signs, and $\nu_q = \pm1$ when $\Lambda$ and $m$ have the opposite signs.  This relation between the topology in momentum space of the fermion dispersion relation and the quantization of the Hall current in 2+1 dimensions has been remarked on previously in connection with the Ward-Takahashi identity in Refs.~\cite{ishikawa1987microscopic,Golterman:1992ub}.

 We arrive at the index of $\CD$ as the surface integral at infinity in Euclidian 3-space
 \beq
 \text{ind}(\CD) &=& -\half \int_S \, \CJ_\alpha dS_\alpha =  - \frac{\nu_q}{2\pi} \int_S \,  \epsilon_{\alpha\beta\gamma} \partial_\beta A_\gamma\,  dS_\alpha \,
 \eeq
 and it remains for us to show that what is multiplying $\nu_q$ is an integer winding number in coordinate space.
For our diagnostic gauge field, we choose $A_{0,1}$ to be independent of $y$, while $A_2=0$, and for our integration region we take the volume to be a cylinder with its axis perpendicular to the domain wall, which we then take to infinite size in every direction.  The surface integral  only gets contributions from the end caps of the cylinder,  $\nu_q$ being different at the two ends when we assume that $\Lambda$ (with $\Lambda>0$) and $m$ have the same sign for $y>0$ and the opposite signs for $y<0$. The expression for the index then becomes
 \beq
  \text{ind}(\CD) &=& - \left(\oint \vec A\cdot d\vec\ell\right) \,\nu_q(y)\biggl\vert^{y=\infty}_{y=-\infty} 
=  -\nu_A \nu_q(y)\biggl\vert^{y=\infty}_{y=-\infty} 
=    -\frac{\nu_A}{2} \left[\frac{\Lambda}{|\Lambda|} - \frac{m(y)}{|m(y)|}\right]^{y=\infty}_{y=-\infty} = \nu_A\ ,\cr &&
 \eqn{india} \eeq
  where $\nu_A$ is the winding number of our Abelian gauge field integrated over the circle at infinity in the $\tau-x$ plane (where $\tau=x_0$ and $x=x_1$),  evaluated at $y = x_2=\pm \infty$.
  We see that a diagnostic gauge field with winding number $\nu_A=1$ yields $\text{ind}(\CD) =1$,  revealing the existence of the gapless edge state in the corresponding Minkowski spacetime theory through a combined topological configuration in both coordinate and momentum space. Again, with more flavors we could get any integer value for the index, consistent with the fact that this model is in the $D$ class (antisymmetric $C$ and broken $T$ symmetry) for which the topological invariant is known to be ${\mathbb Z}$ in two spatial dimensions.

      
 \subsection{A $d=2+1$ Majorana fermion with only ${\mathbb Z}_2$ fermion  number symmetry}
 
\label{sec4}
 The model becomes more interesting when we explicitly break fermion number symmetry down to ${\mathbb Z}_2$ by adding a Majorana mass (with the domain wall profile still in the Dirac mass).  This system is analogous to a topological superconductor with the Majorana mass playing the role of the condensation of Cooper pairs.  As it has no conserve fermion number, there is no conventional Hall current in this model, even though, as we shall see, for some parameters there exist gapless edge states.  The Minkowski theory is
\beq
\CL_\text{M} &=& \bar\psi \left(i\slashed{\partial} - m\right)\psi + \frac{i\mu}{2}\psi^T C \psi +\frac{ i\mu}{2}\bar \psi C\bar \psi^T \ ,\eqn{majmod}
\eeq
where $m$ is real, we can take   $\mu$ to be real and positive, and $C$ is the charge conjugation matrix satisfying
\beq
 C^\dagger = C^{-1} = C\ ,\qquad C\gamma^\mu C^{-1} = -(\gamma^\mu)^T\ .
 \eeq
Once again we will assume a domain wall profile $m(y)$, while keeping the Majorana mass $\mu$ constant. Because of the lack of continuous symmetry in this model, there are no conserved currents and hence no anomaly current in-flow picture at nonzero $\mu$.  Nevertheless, we will show that one can still detect massless edge states in this model by computing the generalized Hall current, which does have in-flow onto the defect when massless edge states exist, and we show that we can use this inflow to count such states.

Since Dirac notation is cumbersome when fermion number is violated, our first step is to rewrite $\CL_\text{M}$ in terms of two real spinor fields $\chi_{1,2}$, where $\psi =( \chi_1+i\chi_2)/\sqrt{2}$.
The Lagrangian can then be expressed in terms of a 4-component spinor 
\beq
\chi =\begin{pmatrix} \chi_1\\ \chi_2\end{pmatrix}
\eeq
as
\beq
\CL_\text{M} &=& \frac{1}{2}\chi^T \left[
\begin{pmatrix*}[r]
1 & \phantom{-}i\\ -i & 1
\end{pmatrix*}
\otimes\gamma_M^0\left(i\slashed{\partial} - m\right)
 - \mu \begin{pmatrix*}[r] 0 &  1 \\ 1 &  0 \end{pmatrix*}\otimes  C
  \right ]_\CA \chi\cr &&
 \eqn{mats} \eeq
  where the subscript ``$\CA$''  means ``antisymmetric part'', derivatives being antisymmetric. We can now Wick rotate to Euclidean space and write the Euclidean Lagrangian as
  \beq
\CL_\text{E} &=& \frac{1}{2}\chi^T \left[
\begin{pmatrix*}[r]
1 & \phantom{-}i\\ -i & 1
\end{pmatrix*}
\otimes\gamma_0\left(\slashed{\partial} +m\right)
 + \mu \begin{pmatrix*}[r] 0 &  1 \\ 1 &  0 \end{pmatrix*} 
 \otimes C \right ]_\CA \chi\cr &&
 \eqn{matsE} \eeq
 In the particular $\gamma$-matrix basis of \eq{d3basis}
 \beq
\gamma_0 =C=\sigma_2 \ ,\qquad
\gamma_1 =- \sigma_1\ ,\qquad
\gamma_2 =\sigma_3 \ ,
\eeq
we can write $\CL_\text{E}$ 
in terms of the fermion fields
\beq
\zeta= \frac{1}{\sqrt{2}}\begin{pmatrix}\chi_1+\chi_2 \\ \chi_1-\chi_2\end{pmatrix} \equiv \begin{pmatrix} \zeta_+ \\ \zeta_- \end{pmatrix}
\eeq
as
 \beq
 \CL_E = \half\left[\zeta_+^T C \CD_+ \zeta_+ +\zeta_-^T C \CD_- \zeta_- \right]\ , \qquad \CD_\pm = \slashed{\partial} + (m\pm \mu)\ .
\eqn{D2}
 \eeq
 
 As in the $1+1$ dimensional case we can drop the $C$ matrix and use $[\text{ind}(\CD_+) + \text{ind}(\CD_-)]$ to detect edge states in this case, with $m(y)$ having a domain wall form and $\mu$ being constant.  We can immediately adapt the result  \eq{india} in the previous section and to evaluate this index as
  \beq
\text{ind}(\CD)\equiv  [\text{ind}(\CD_+) + \text{ind}(\CD_-)]  &=&  -\left(\oint \vec A\cdot d\vec\ell\right) \,\left(\nu^+_q(y)+\nu^-_q(y)\right)\biggl\vert^{y=\infty}_{y=-\infty} \cr
&=&  -\nu_A \left(\nu^+_q(y)+\nu^-_q(y)\right)\biggl\vert^{y=\infty}_{y=-\infty} \ ,\cr &&
\eeq

 where
 \beq
 \nu_q^{(\pm)}  =\half \left(\frac{\Lambda}{|\Lambda|} - \frac{m\pm\mu}{|m\pm\mu|}\right)\ .
 \eeq
Assuming positive $\Lambda$, we have
 \beq
 \left(\nu_q^{(+)} + \nu_q^{(-)}\right) = \begin{cases} 
 2 & m<-|\mu|\ ,\\
 1 & -|\mu| < m < |\mu| \ ,\\
 0 & |\mu| < m\ . 
 \end{cases} 
 \eqn{case}
 \eeq
 The analog of \eq{india} then follows for the index in this theory, 
  \beq
  \text{ind}(\CD) &=&-\nu_A\,\left[\nu^{(+)}_q(y)+\nu^{(-)}_q(y)\right]\bigl\vert^{y=\infty}_{y=-\infty} \ ,
\eqn{indiamu} 
\eeq
where $\nu_A$ is the winding number in the diagnostic gauge field.  For $\nu_A=1$ it follows that the index takes on one of the values $0,\pm1,\pm2$  depending on which the cases in \eq{case} pertains for the asymptotic values of  $m$ at the two sides of the domain wall.  If we denote the asymptotic values $m(\pm\infty) = m_\pm$, then our result for the index for various cases is given  in Table~\ref{tab:index3}.

\begin{table}[t]
  \begin{center}
    \begin{tabular}{c|c|c|c|} 
  &$\  m_+<-|\mu| \ $&$\  -|\mu| < m_+<|\mu| \ $&$\  |\mu|<m_+\  $\\
 \hline $m_-<-|\mu| \ $&$ \phantom{-} 0 $&$  \phantom{-}  1 $&$ \phantom{-} 2$\\
\hline $ -|\mu| < m_-<|\mu|\  $&$ -1 $&$ \phantom{-} 0 $&$ \phantom{-} 1$\\
 \hline  $ |\mu|<m_- \ $&$ -2 $&$ -1 $&$  \phantom{-}0 $\\
    \end{tabular}
    \caption{The index $\text{ind}(\CD)$ for spatial topology $\nu_A=1$ as a function of $m_\pm$, the asymptotic values of the fermion mass on the two sides of the domain wall, relative to the constant Majorana mass $\mu$.}
    \label{tab:index3}
  \end{center}
\end{table}

We now show that the above results agree with what one finds when constructing explicit domain wall solutions for this model.  We return the Minkowski Lagrangian
\beq
\CL_M = \bar\psi \left(i\slashed{\partial} - m\right)\psi + i\frac{\mu}{2}\psi^T C \psi + i\frac{\mu}{2}\bar \psi C\bar \psi^T \ ,
\text{sign of }\mu
\eqn{majmod}\eeq
and consider the equation of motion in our basis
\beq
\gamma^0 = \sigma_2\ ,\quad \gamma^1=-i\sigma_1\ ,\quad \gamma^2=i\sigma_3\ ,\quad C = \sigma_2\ .
\eeq
To obtain the  solutions to the equations of motion we redefine the fermion field $\psi=e^{i\frac{\pi}{4}}\varphi$
in which case we can write the solutions as 
\beq
\varphi(x)=
\begin{pmatrix}
\alpha_1 e^{(-m-\mu ) y}+i\beta_1 e^{(-m+\mu)y}\\
\alpha_2 e^{(m+\mu)y}+i\beta_2 e^{(m-\mu)y}
\end{pmatrix}
\eqn{ysols}
\eeq 
We take $\mu$  to be spatially constant, and without a loss of generality we assume $\mu>0$, while the mass $m$ depends on the coordinate $y$ and takes the asymptotic values $m\to m_\pm$ as $y\to \pm \infty$.   For $y >0$, a localized solution requires that the coefficient of $y$ in the exponent  in \eq{ysols} be negative when $m$ is replaced by $m_+$.    Out of the four exponentials in \eq{ysols}, there are always two that meet this criterion.  Labeling them by their coefficients, they are
\begin{equation}
\begin{aligned}
m_+ > |\mu|:&\qquad  &\alpha_1,\,\beta_1\ ,&\\
-|\mu|< m_+ < |\mu|: & &\alpha_1,\,\beta_2 \ ,&\\ 
m_+ < -|\mu|:&\qquad  &\alpha_2,\,\beta_2\ .&
\end{aligned}
\end{equation}

We can do the same thing for solutions at $y<0$ and $m\to m_-$; in this case the coefficient of $y$ must be positive, and again there are always two solutions, the same as the above but with subscripts $1,2$ reversed:
\begin{equation}
\begin{aligned}
m_- > |\mu|:&\qquad  &\alpha_2,\,\beta_2\ ,&\\
-|\mu|< m_- < |\mu|: &&\alpha_2,\,\beta_1 \ ,&\\
m_- < -|\mu|:&\qquad  &\alpha_1,\,\beta_1\ .&
\end{aligned}
\end{equation}

 When we match solutions at $y=0$ there must localized solutions of the same chirality on both sides, and they must be both real or both imaginary.  That means that the must be the same $\alpha_i$ or $\beta_i$ solution for positive and negative $y$.  It is evident then that depending on the two values $m_\pm$ relative to $|\mu|$ there can be 2, 1, or 0 solutions.  For example, if   $m_+>|\mu|$ while   $-|\mu|<m_- < |\mu|$, then there is one localized positive chirality $\beta_1$   solution that can be matched across $y=0$.  The number and chirality of edge state solutions are given in Table~\ref{tab:DWF3}, where the $R,L$ entry tells us whether we have upper or lower component solutions respectively.

  \begin{table}[t]
  \begin{center}
    \begin{tabular}{c|c|c|c|} 
 &$\  m_+<-|\mu| \ $&$\  -|\mu| < m_+<|\mu| \ $&$\  |\mu|<m_+\  $\\
\hline $m_-<-|\mu| \ $&$  0 $& R& 2R\\
\hline  $-|\mu| < m_-<|\mu|\  $& L&$0 $& R\\
 \hline  $ |\mu|<m_- \ $& 2L& L &$ 0 $\\\\
    \end{tabular}
    \caption{ The number and chirality of edge state solutions to $\CD\psi=0$ before introducing a gauge field,  where the R,L indicates chirality. The table for solutions to $\CD^\dagger\psi=0$ would be the same with substitution $\text{L}\leftrightarrow \text{R}$.}
    \label{tab:DWF3}
  \end{center}
\end{table}
  Note that the index in Table~\ref{tab:index3} is giving us the number of positive chirality massless edge states minus the number of negative chirality ones.  The reason for that is that  the equation $\CD^\dagger\psi=0$ has the   same solutions as  $\CD\psi=0$ except for a parity flip, exchanging $R\leftrightarrow L$.  When the gauge field is turned on with $\nu_A=1$, then it localizes the $R$ solutions of $\CD$ while delocalizing the $L$ solutions.  Therefore what the index is counting is the number of $R$ solutions for $\CD\psi=0$, minus the number of $R$ solutions for $\CD^\dagger\psi=0$, which is equivalent to the number of positive chirality massless edge states minus the number of negative chirality ones for the operator $\CD$.

    The index we computed still takes values in ${\mathbb Z}$, which is appropriate since the Majorana mass $\mu$ does not break $T$ symmetry and the system remains in the $D$ topological class.


\section{Dirac fermion in $3+1$ dimensions}

Our last example of a Dirac fermion $3+1$ dimensions shares many features with the $1+1$ dimension example. When the mass $m$ has a domain wall profile, the Dirac equation  in Minkowski spacetime has exact solutions corresponding to 2-component massless edge states localized on the $2+1$ dimensional wall. Such a domain wall describes the physics of a topological insulator, where the region with $m<0$ is considered the interior of the topological insulator  and the region with $m>0$ is considered to be the exterior.  
There is no analog of chirality in $2+1$ dimensions, and hence no charge violation on the wall in the presence of background gauge fields and no  inflowing current from the bulk to the wall maintaining current conservation. This makes the theory of three dimensional topological insulator another interesting example to which one could imagine applying our construction for which to compute the generalized Hall current and index.

The Euclidean Dirac operator is simply $\slashed{\partial} + m(x_3)$ which has a static and unnormalizable solution localized at $x_3=0$ and constant in all the other coordinates.  The edge state has two nonzero spinor components and is an eigenstate of  $\gamma_3$.  To fully localize this state we add as diagnostic fields a four dimensional gauge field and a pseudoscalar, and again we consider the mass to be an arbitrary scalar field for now.   Thus the operator we consider is
\beq	
\CD = D_\mu \gamma_\mu+ \phi_1+i \phi_2\gamma_\chi\ ,
\eqn{D}
\eeq
where  $D_\mu = (\partial_\mu + i A_\mu)$ is the $d=4$ gauge covariant derivative and $\gamma_\mu$, $\gamma_\chi$ are our 4d Dirac matrices.

To compute the index of $\CD$ we once again construct the $K$ operator,
\beq
K= \begin{pmatrix} 0 & -\CD^\dagger \\ \CD & 0\end{pmatrix} = D_\mu \Gamma_\mu + i\phi_2 \Gamma_4 + i \phi_1 \Gamma_5\ ,
\eeq
where $\mu = 0,\ldots 3$ and 
the $\Gamma_a$ are the 8-dimensional matrices
\beq
\Gamma_\mu &=&\sigma_1\otimes \gamma_\mu\ ,\qquad \mu = 0,\ldots,3\ , \cr
\Gamma_4 &=& \sigma_1\otimes \gamma_\chi\ ,\cr
\Gamma_5 &=& -\sigma_2\otimes 1\ ,\cr
\Gamma_\chi &=& \sigma_3\otimes 1\ .
\ \text{index starts at 0}\eeq

Our task is to compute the part of the chiral current $\CJ_a = \bar\Psi \Gamma_a \Gamma_\chi \Psi$ that contributes to the index, where $\Psi$ is a fermion with action $S = \bar\Psi K \Psi$.  As in the $1+1$ dimensional example in \eq{scalar}, we first write the scalars as
\beq
\phi &=& \phi_1 + i \phi_2 = (v+\rho(x)) e^{i\theta(x)}
\eqn{scalar2}
\eeq
assuming  constant $v$ and slowly varying $\rho$ and $\theta$ where $\rho=\theta = 0$. We compute the leading contribution to the chiral current in a $1/v$ expansion, since higher order terms will drop off too fast at infinity to contribute to the integral $\int \partial_\mu \CJ_\mu$.  To this end we write
\beq
K &=& K_0 + \delta K\ ,
\eeq
with
\beq
 K_0 = 
 \partial_\mu \Gamma_\mu + i  v \Gamma_5\ ,\qquad
\delta K = i A_\mu \Gamma_\mu + i\theta   v \Gamma_4 + i  \rho(x)  \Gamma_5   \  .
\eeq

Then $K_0^{-1}$ will be the free fermion propagator, while we perturb in $\delta K$. To compute the part of the current that contributes to the index we need the leading term in a $1/v$ expansion, since higher order terms will drop off too fast at infinity to contribute to the integral $\int \partial_\mu \CJ_\mu$.

\begin{figure*}[t]
\centerline{\includegraphics[height=3 cm]{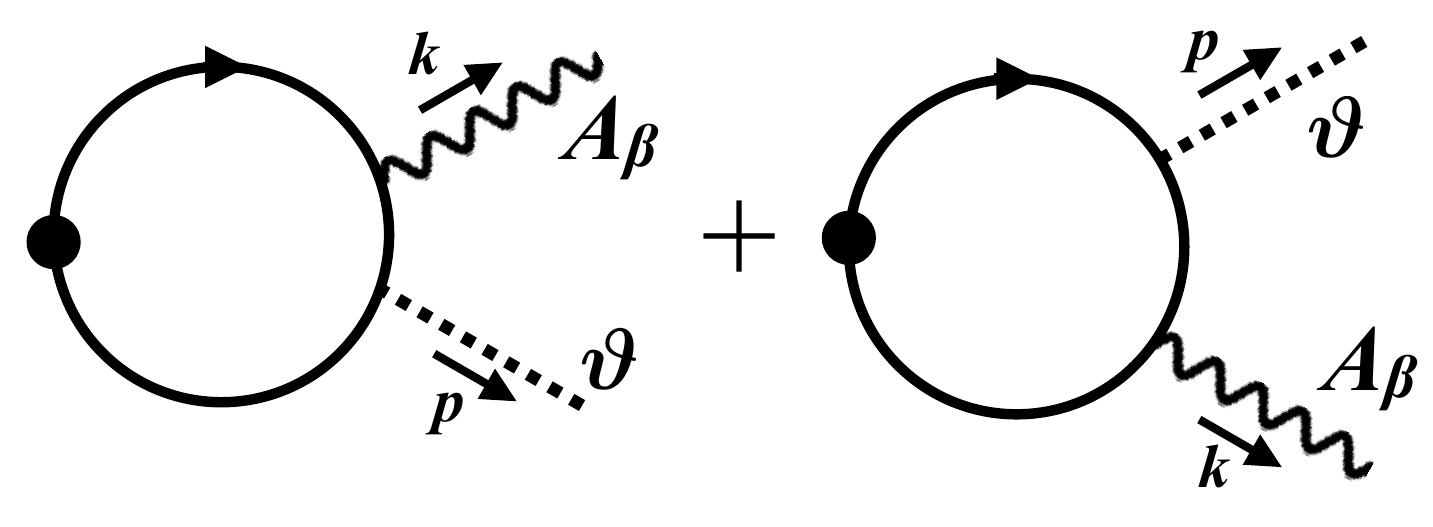}}
\caption{\it Loop diagrams for computing the generalized Hall current for the   $3+1$-dimension Dirac fermion. The black dot is an insertion of the chiral current $\Gamma_\alpha\Gamma_\chi$ with incoming momentum $p+k$. The outgoing fields are the gauge field and the phase $\theta$ of the complex field $\phi_1+i\phi_2$, and the fermion propagator is given by $K_0^{-1}$.}
\label{fig:4dloop}
\end{figure*}

When expanding $\CJ_\mu$  in $\delta K$,  the $\Gamma_\chi$ insertion in the fermion loop requires that the rest of the graph supplies one each of the other six $\Gamma_a$ matrices in order to get a nonzero contribution from the trace.  First consider the source of $\Gamma_{4,5}$ in the graph.   We see the $\Gamma_5$ can come from one of the fermion propagators $K_0^{-1}$, but to obtain $\Gamma_4$ we require an insertion of $\theta$ in the graph, while to lowest order the $\rho$ contribution will vanish.  To obtain the other four $\Gamma_\mu$  we note that the result for $\CJ_\mu$ will be proportional to an epsilon tensor $\epsilon_{\mu\alpha\beta\gamma}$ and that we can only contract the $\alpha,\beta,\gamma$ indices with a gauge field and two derivatives -- one acting on the gauge field, the other on $\theta$.  Thus we must expand the graphs in  Fig.~\ref{fig:4dloop} to linear order in the momenta carried by the gauge field and by $\theta$.  

It is straightforward to  evaluate this loop integral, with the result
\beq
\CJ_\mu&=&
v \partial_\gamma \theta\partial_\alpha A_\beta
\int \frac{d^4q}{(2\pi)^4}\,
\Tr\Bigl(\Gamma_\mu \Gamma_\chi \left[(\partial_\alpha \tilde K_0^{-1} )\Gamma_\beta  \tilde K_0^{-1} \Gamma_4\left(\partial_\gamma \tilde K_0^{-1} \right)\right. \cr &&
+ \left.(\partial_\gamma \tilde K_0^{-1} )\Gamma_4  \tilde K_0^{-1} \Gamma_\beta\left(\partial_\alpha \tilde K_0^{-1} \right)\right]\Bigl)\cr
&=&
-\epsilon_{\mu \alpha\beta\gamma}\partial_\gamma \theta \partial_\alpha A_\beta\int \frac{d^4q}{(2\pi)^4}\frac{16 v^2}{(q^2+v^2)^3}
\cr &=&
-\frac{1}{2\pi^2} \epsilon_{\mu\alpha\beta\gamma}\partial_\gamma \theta \partial_\alpha A_\beta\ ,
\eqn{J31b}\eeq
where the derivatives inside the integral are with respect to $q$.  The prefactor in the first line includes a $(-1)$ for the fermion loop, $(-1)$ from the Fourier transform $p^2\to -\partial_x^2$, $(-1)$ from the two derivatives with respect to momentum with opposite signs due to momentum flow, and factors $(-iv)$ and $-i\Gamma_\beta$ for $\theta$ and $A_\beta$ vertices respectively.   

To highlight the momentum space topology underlying the above calculation, we follow similar steps as in the $1+1$ dimensional example of \S~\ref{sec:3}  to rewrite the result as the analog of \eq{D3new}
\beq
\CJ_\mu&=&
\frac{1}{12}\epsilon_{\mu\alpha\beta\gamma} \partial_\gamma \theta\partial_\alpha A_\beta
\cr &&\times\epsilon_{\rho\sigma\tau\omega}
\int \frac{d^4q}{(2\pi)^4}\,
\Tr\gamma_\chi\left[\left( \xi^\dagger\partial_\rho\xi\right)\left( \xi^\dagger\partial_\sigma\xi\right)\left( \xi^\dagger\partial_\tau\xi\right)\left( \xi^\dagger\partial_\omega\xi\right)+\left(\xi\partial_\rho\xi^\dagger\right)\left( \xi\partial_\sigma\xi^\dagger\right)\left(\xi\partial_\tau\xi^\dagger\right)\left( \xi\partial_\omega\xi^\dagger\right) \right]\cr &&
\eqn{D4new}
\eeq
where
\beq
\xi = \frac{\tilde\CD_0}{(\det \tilde\CD_0)^\frac{1}{4} }= \frac{v}{\sqrt{q^2+v^2}} + i \frac{q_\mu \gamma_\mu}{\sqrt{q^2+v^2}} \ .
\eeq
Making use of the currents $A_\mu$ and $V_\mu$ defined as
\beq
A_\mu = \frac{i}{2}\left(\xi^\dagger\partial_\mu \xi - \xi \partial_\mu \xi^\dagger\right)\ ,\qquad V_\mu =  \frac{1}{2}\left(\xi^\dagger\partial_\mu \xi +\xi \partial_\mu \xi^\dagger\right)\ ,
\eqn{AVdef}
\eeq
the integrand  in \eq{D4new}  can be related to the volume form derived for four spacetime dimensions in \eq{rtg} using the relation that one can derive
\beq
&&\hspace{-.2in}-\frac{\epsilon_{\rho\sigma\tau\omega}}{96\pi^2} \left[\left( \xi^\dagger\partial_\rho\xi\right)\left( \xi^\dagger\partial_\sigma\xi\right)\left( \xi^\dagger\partial_\tau\xi\right)\left( \xi^\dagger\partial_\omega\xi\right)+\left(\xi\partial_\rho\xi^\dagger\right)\left( \xi\partial_\sigma\xi^\dagger\right)\left(\xi\partial_\tau\xi^\dagger\right)\left( \xi\partial_\omega\xi^\dagger\right) \right] \cr \qquad
&=&\epsilon_{\rho\sigma\tau\omega}\left[-\frac{1}{16\pi^2}   A_\rho A_\sigma A_\tau A_\omega  + \partial_\sigma  \left( \frac{1}{24\pi^2} A_\sigma A_\tau V_\omega\right)\right]\ .
\eeq
On taking the trace of the above equation multiplied by $\gamma_\chi$, the first term on the right is recognized from \eq{rtg} as the volume measure for $S^4$, while the second term is the divergence of a current which is well defined without singularities on $S^4$, and hence integrates to zero.  Therefore the generalized Hall current in \eq{D4new} can be written as
\beq
\CJ_\mu&=&-\frac{\nu_q}{2\pi^2} \epsilon_{\mu\alpha\beta\gamma}\partial_\gamma \theta \partial_\alpha A_\beta\ ,
\eqn{J31c}
\eeq
where $\nu_q$ is the element of $\pi_4(S^4)$:
\beq
\nu_q = -\frac{1}{16\pi^2}\epsilon_{abcd}\int d^4 q \,  \Tr \gamma_\chi A_a A_b A_c A_d  = 1\ .
\eeq

The index is  given by
\beq
\text{ind}(\CD) =-\half \int d^4x \partial_\mu \CJ_\mu = 
\frac{\nu_q}{4\pi^2} \int _S \epsilon_{\mu \alpha\beta\gamma} \partial_\gamma \theta \partial_\alpha A_\beta \,dS_\mu\ .
\eqn{J31d}
\eeq
To evaluate the spatial integral we first must choose a topologically nontrivial set of diagnostic fields to localize the solutions to $\CD\psi=0$.  That can be done by making use of  the BPS monopole field configuration discussed in Ref.~\cite{cheng2013fermion} which considers a massless Dirac fermion in three Euclidean dimensions interacting with a scalar field and a gauge field,
\beq
\Phi = \frac{g}{2\pi}\left(a-\frac{1}{2r}\right)\ ,\qquad {\bf A} = -\frac{g(1+\cos\theta)}{4\pi r \sin\theta}\,\hat {\bf e}_\varphi\ ,
\eqn{CF} 
\eeq
 and discuss how  when the couplings obey the minimal Dirac quantization relation, $eg = 2\pi$, the $d=3$ conjugate Dirac operator $\CD_3^\dagger$  has a zeromode proportional to $\exp(-a r)$ while $\CD_3$ has none.  This configuration can be lifted into our four-dimensional Euclidean theory by  taking   $ {\bf A} $ to be the first three components of our four-component gauge field, independent of the fourth coordinate, while
\beq
\phi=m(x_4)+i \Phi(\bfx)\ ,\qquad A_4=0\ ,\qquad F_{ij} =  \epsilon_{ijk4} \frac{g \hat r_k}{4\pi r^2}\ ,
\eeq
where for notational convenience we have relabeled our coordinates so that the domain wall mass is a function of $x_4$ and coordinates on the mass defect are labeled by $x_{1,2,3}$. Since we have set the electric charge $e=1$ in our covariant derivative, we take $g=2\pi$ for the magnetic charge.
With these background fields  we then expect that the existence of a massless edge state in the original $d=3+1$ Minkowski theory implies one zeromode for $\CD^\dagger$ only if $a>0$ and no zeromode for $\CD$, so that $\text{ind}(\CD)=-\theta(a)$.

This is indeed what we find. Plugging in these fields into \eq{J31c}, we get the generalized Hall current
\beq
\CJ_i &=&\frac{\hat \bfr_i}{2\pi^2 r} \frac{(2ar-1)\frac{{\rm d}m(x_3)}{{\rm d}x_3}}{(2ar-1)^2 + 4 r^2 m(x_3)^2}\ ,\quad \CJ_3 = \frac{m(x_3)}
{2\pi^2 r^2\left((2a r-1)^2 + 4 r^2 m(x_3)^2\right)}\ ,\cr &&
 \eqn{J31e}
 \eeq
 for $i=0,1,2$.
We then compute the integral of its divergence,
\beq
\int d^4x\, \partial_\mu J_\mu = \lim_{L_3,R\to\infty}\left[\int^R d^3 r \,J_4\biggl\vert^{x_3=L_3}_{x_4=-L_3} + \int_{-L_3}^{L_3} dx_3 \int d\Omega \,R^2\,\hat R\cdot \vec J\right]\ .
\eeq
Specializing to the mass profile $m(\pm\infty) = \pm m_0$, 
and using the expression for current given in \eq{J31e} we find
\beq
 \lim_{L_3,R\to\infty} \int_{-L_4}^{L_4} dx_4 \int d\Omega \,R^2\,\hat r\cdot \vec J &=&
 \frac{2 \tan ^{-1}\left(\frac{ m_0}{a}\right)}{\pi }\ ,\cr &&\cr
  \lim_{L_3,R\to\infty}\int^R d^3 r \,J_3\biggl\vert^{x_3=L_3}_{x_3=-L_3} &=& \left(1 + \frac{2\tan^{-1} \frac{a}{m_0}}{\pi}\right)
 \eeq
Summing these two terms gives us
\beq
\int d^4x \, \partial_\mu \CJ_\mu = 2\theta(a)
\eeq
and yields the anticipated result for the index,
\beq
\text{ind}(\CD) =-\half \int d^4x \, \partial_\mu \CJ_\mu = -\theta(a)\ .
\eeq
So we see again that with diagnostic fields to localize massless edge states, the index of the Euclidean fermion operator counts these edge states and is quantized because of both the spacetime topology of the background fields, and the momentum space topology of the fermion dispersion relation. As in the $d=1+1$ case and unlike the $d=2+1$ example, we find that the $d=3+1$ result is not affected by regulator fields, which decouple.

This model is in the DIII class, symmetric under both $C$ and $T$ with both $C$ and $T$ matrices antisymmetric  \cite{PhysRevB.78.195125}.  The topological invariant in this case for three spatial dimensions is ${\mathbb Z}$, consistent with what we would find if we generalized this theory to more flavors without any flavor symmetry.

\section{Interactions}
\label{sec:interactions}

Our interest in understanding edge states via the index of the Euclidean fermion operator led us to computing the incoming flux at infinity of the generalized Hall current.  This has all been for free fermions, but interacting systems are more interesting \cite{PhysRevB.85.245132, PhysRevB.88.064507, Gu:2013azn, PhysRevB.92.085114, Gu:2015wli, PhysRevB.90.245120, PhysRevB.95.195108, PhysRevLett.117.206405, PhysRevB.92.085114, Wang:2022ucy}, and  we know that interactions can change the topological classification \cite{Fidkowski:2009dba}.  The ramifications go beyond condensed matter systems and have been applied to lattice models for chiral gauge theories, where vector-like fermions appear as chiral edge states until interactions are turned on, gapping some of them and leaving behind a chiral representation \cite{Wang:2013yta,You:2014vea,Wang:2018ugf,Zeng:2022grc}. For an analogous discussion in continuum chiral gauge theories, see \cite {Tong:2021phe, PhysRevX.11.011063}. Clearly the index of the free fermion operator cannot capture this physics. The generalized Hall current, on the other hand, is well-defined even in the presence of interactions, and so it is reasonable to ask whether its divergence still tells us about massless edge states in an interacting theory.  In this section we speculate that that is plausible, in the context of the same $1+1$ model described in Ref.~\cite{Fidkowski:2009dba}.

First we examine more closely how the calculations for free fermions were done.  Our method followed the work of Callan and Harvey \cite{Callan:1984sa}, which in turn used the methods developed by Goldstone and Wilczek \cite{Goldstone:1981kk}.  We computed the generalized Hall current in a derivative expansion, integrating out the fermions in a background field.  In the $d=1+1$ example, this background field was a complex scalar field, and we obtained a contribution to the current proportional to $\partial_\mu\theta = i\phi^*\overleftrightarrow {\partial_\mu} \phi/|\phi|^2$.  Recall that the index was defined in  \eq{indform} as the integral of the divergence of the generalized Hall current in the limit that the doubled fermion's mass $M$ tended to zero.  As seen in \eq{inddef}, the mass $M$ was introduced as an infrared cutoff, and in the Ward-Takahashi identity for the current in \eq{WI},
\beq
\partial^\mu \CJ_\mu^\chi = 2M\bar\Psi\Gamma_\chi\Psi - \CA\ ,
\eeq
where $\CA=0$. The nonzero divergence indicating the existence of the mass less edge state comes from the $2M\bar\Psi\Gamma_\chi\Psi $ term on the right hand side, in the limit that $M\to 0$, where $M$ serves as an IR regulator.  In our calculation, the inverse dependence on $|\phi|^2$ is the sign of an infrared divergence regulated by $\vev{\phi}$ -- it was because the background field served as an IR regulator that we could set the parameter $M$ in  \eq{indform} to zero before computing the Feynman diagrams. We replaced $M$ by a spatially varying $\phi$ field as the IR regulator so that a nonzero index of the fermion operator $\CD$ would indicate there was a massless edge state when $\phi$ was removed. The calculation was performed as if the fermions were fully gapped by by $\phi$.  This is clearly false, since we were studying systems with an exact zeromode.  The current we computed cannot be valid in the region where the zeromode wave function is appreciable since the fermion is not gapped there and the derivative expansion in the background field breaks down.  However, since the index is proportional to $\int \partial_\mu\CJ_\mu$,  it only depends on the current asymptotically far away from the localized zeromode,  For that reason we were able to treat the fermion as gapped, and justify the derivative expansion. This is the same justification as for the Callan-Harvey calculation \cite{Callan:1984sa}. 

We will argue that in theories where interactions fully gap the fermions, this spatially constant gap will serve as the dominant infrared regulator in calculating the  generalized Hall current, and since it cannot localize the fermions, it will lead to  a vanishing divergence of the current.  We envision a mechanism similar to what we saw in  \S~\ref{sec:3} where we  discussed the example of $2N$ flavors of free, $d=1+1$ Majorana fermions in the presence of time reversal violation.  There we found that if we explicitly broke time reversal invariance via a spatially constant $i\mu \psi^T C\gamma_\chi\psi$ term, the index vanished.  The reason was that  in this case, as we removed the diagnostic field  $\phi_2(x)\to 0$, the spatial topology experienced by the fermion abruptly became trivial at the point where $\mu$ dominated over $\phi_2$  as the infrared regulator.
The question then is whether the same phenomenon can occur when interactions gap the fermions  as we remove the diagnostic fields.  This seems plausible, making the procedure described here for detecting massless edge states relevant when interactions are introduced, even though the original motivation of looking at zeromodes of the free fermion operator is no longer applicable.  To understand this better, we look at the model of Ref.~\cite{Fidkowski:2009dba} in greater detail.

Consider $2N$ copies of the $1+1$ dimension Majorana fermion model discussed in \S~\ref{sec:3}. As shown in \eq{d2DWF} the Minkowski domain wall solutions in the free theory with a step function mass  take the form
\beq
\psi_i(x,t)  =\eta_i(t) e^{-m|x|} \begin{pmatrix}\phantom{-} 1 \\ -1\end{pmatrix}\ ,
\eeq
and when quantized, the hermitian $\hat \eta_i$ operators obey the Clifford algebra
\beq
\{\hat \eta_i,\hat \eta_j\} = 2\delta_{ij}\ ,\qquad i = 1,\ldots,2N.
\eeq
From these we can construct the ladder operators
\beq
\hat c_a = \frac{\hat \eta_a + i \hat \eta_{a+N}}{2} \ ,\qquad \hat c^\dagger_a =  \frac{\hat \eta_a - i \hat \eta_{a+N}}{2}\ ,\qquad a=1,\ldots N\ ,
\eqn{cdef}\eeq
which obey the usual fermion anticommutation relations
\beq
\{\hat c_a,\hat c_b\} = \{\hat c^\dagger_a,\hat c^\dagger_b\} = 0\ ,\qquad \{\hat c_a, \hat c^\dagger_b\} = \delta_{ab}\ .
\eeq
The $2^{N}$-fold degenerate edge states can then be constructed by acting with $c_a^\dagger$ operators on a state $\ket{0}$, which is defined to be the state annihilated by all of the $\hat c_a$ operators. 

Since the $\hat \eta_i$ operators obey a Clifford algebra, they define an $\mathfrak{so}$(2N) Lie algebra, under which the degenerate ground states transform as the $2^N$ dimensional  reducible spinor representation.  When considering interactions between the edge states, it is convenient to represent the $\hat \eta_i$ operators as  $2^N\times 2^N$ hermitian Dirac gamma matrices which act on these states.  Interactions between these states  can then be represented as a  matrix consisting of sums  of  totally antisymmetrized products of even numbers of gamma matrices, which we will call $H_\text{int}$.  

One constraint we will impose on the interactions is that they preserve time reversal symmetry, since the  gapless edge states in the free theory owe their existence to that symmetry in the first place, as discussed in \S~\ref{sec:3Nf}.  The action of time reversal on the bulk states is $\psi\to \sigma_1\psi$ which takes $\hat \eta_i\to -\hat \eta_i$.  This sign is not interesting  since we will only be considering products of an even number of the $\hat \eta_i$ operators; however, in order for an interaction represented an antisymmetrized product of $2k$ antisymmetrized gamma matrices to be hermitian it must be proportional to $i^{k}$, which means that operators with odd $k$ flip sign under the antiunitary time reversal transformation. So we restrict the interaction to operators involving products of multiples of {\it four} fermion fields.

One of the results of Ref.~\cite{Fidkowski:2009dba} is that this time reversal invariant $H_\text{int}$ can gap all of the edge states in this model if and only if the number of flavors is a multiple of eight. Completely gapping the edge states means that there is a unique, nondegenerate ground state -- so to prove this result we need to show that a sum of totally antisymmetrized products of $4k$ $SO(2N)$ gamma matrices can only have a unique lowest eigenvalue   when $2N = 0\mod 8$.  This is easy to show and the argument is given in Appendix~\ref{sec:appendix_mod8}.  

Consider the case with eight flavors of fermions and consider the interaction defined in \eq{Hint8} and \eq{Lgap}.  On the domain wall,
\beq
H_\text{int} = \omega\left(\hat c_1\hat c_2\hat c_3\hat c_4 + h.c + 1\right) 
\eeq
which for $\omega>0$ has the unique groundstate $\ket{\Omega} = (\ket{0000} - \ket{1111})/\sqrt{2}$ with energy $E=0$, fourteen degenerate states with $E=\omega$, and an isolated state with energy $E=2\omega$. 
In the effective Euclidean  $0+1$ dimension theory one can compute the 1-particle propagator and find
\beq
\expect{\Omega}{\eta_i e^{-\hat H_\text{int} \tau} \eta_j }{\Omega} = e^{-\omega\tau} \delta_{ij}\ .
\eeq

How do these interaction affect the actual calculation of the divergence of the generalized Hall current in the doubled Euclidean version of the $1+1$ dimension Minkowski theory?    We do not have a quantitative answer but believe that the gapping of the single particle propagator in the fully interacting theory will render the generalized Hall current divergenceless\footnote{For work on the Greens function for theories with gapped edge states, see Refs.~\cite{you2014topological,catterall2017novel,you2018symmetric,xu2021green}.}.  

In the $1+1$ dimension bulk theory we expect   the interactions act like the 't Hooft operator induced by instantons in Ref.~\cite{t1976symmetry}, saturating the zeromodes and serving as topologically trivial infrared cutoff.  In the absence of interactions and diagnostic fields, the fermion propagator we use to compute the generalized Hall current in the $N_f=8$ model defined as
\beq
\frac{
\int\prod_{i=1}^8 d\Psi_i\, e^{-S} \, \Psi_j(y_1) \bar\Psi_k(y_2)}{\int\prod_{i=1}^8 d\Psi_i\, e^{-S}}\ ,\qquad S= \int d^2x \bar\Psi \CD\Psi\ ,
\eqn{prop}
\eeq
is divergent since the denominator vanishes due to the integration over zeromodes of $\CD$ which do not appear in the action.  

Suppose we now add a fully gapping interaction, such as the one given in \eq{Lgap},
\beq
\CL_\text{int} = \omega \left[ \left( \psi_1 + i \psi_5\right)^T C \gamma_\chi \left( \psi_2 + i \psi_6\right)\right]\left[ \left( \psi_3 + i \psi_7\right)^T C \gamma_\chi \left( \psi_4 + i \psi_8\right)\right] + h.c.\ .
\eqn{Lgap2}
\eeq
The natural extension to add in the doubled Euclidian theory is then
\beq
\CL_\text{int} = \omega \left[ \left(\bar \Psi_1 + i\bar \Psi_5\right)\Gamma_3 \left( \Psi_2 + i \Psi_6\right)\right]\left[ \left( \bar\Psi_3 + i \bar\Psi_7\right) \Gamma_3\left( \Psi_4 + i \Psi_8\right)\right] + h.c.\ ,
\eqn{Lgap3}
\eeq
where $\Gamma_3=\sigma_1\times \gamma_\chi$ is the doubled version of $\gamma_\chi$ given in \eq{Gamdef2}.  When this term is added to the action in \eq{prop} the fermion propagator is no longer IR divergent since the zeromodes of $\CD$ now appear in the action.  As a result, one should still find a well defined index as the diagnostic fields are removed -- but as the infrared cutoff arising from the interaction presumably has  trivial spatial topology, we expect  a topological phase transition to a trivial phase with vanishing divergence for the generalized Hall current, similar to that seen in \S~\ref{sec:3Nf}. It would be interesting to develop a quantitative method to perform the calculation in the presence of interactions, but is beyond the scope of this paper.

\section{Discussion}
 We have shown that the presence of the massless edge states in topological matter manifests itself by the inflow of a current --- which we call a generalized Hall current --- in a related system in Euclidian spacetime. The divergence of this current indicates the existence of massless edge states just as the Hall current inflow does for the Integer Quantum Hall Effect. But this current appears in all topological classes in Minkowski spacetime, including those that do not have conserved currents because of a lack of continuous symmetries (such as a topological superconductor), or whose edges do not suffer from chiral anomalies because they lack chiral symmetry (such as topological insulators in $1+1$ and $3+1$ dimensions).  In this sense one arrives at a unified picture for disparate manifestations of topological matter.  

Furthermore, while the original motivation was to study the index of the free Euclidean Dirac operator, the generalized Hall currents can be computed for interacting systems as well, and we gave qualitative arguments for why we expect the utility of such currents to persist.  Whether  this idea can be put on a firmer foundation is an open question.  It is an attractive proposition to  able to compute analytically how interactions affect topological properties in different systems in various dimensions.  If a general theory for gapping massless chiral edge states in $3+1$ dimensions can be derived, that might shed light on what restrictions there are on the matter content of chiral gauge theories regulated on  a lattice; this is of obvious interest given that the Standard Model is a chiral gauge theory. 

 \section{Acknowledgements}
We thank Lukasz Fidkowski and Michael Clancy for useful conversations. 
DBK is supported in part by DOE Grant No. DE-FG02-00ER41132 and  by the DOE QuantISED program through the
theory  consortium ``Intersections of QIS and Theoretical Particle
Physics'' at Fermilab. SS acknowledges support from the U.S. Department of Energy,
Nuclear Physics Quantum Horizons program through the Early Career Award DE-SC0021892

\appendix

\section{Topology in momentum space for Dirac operators}
\label{sec:appendix_topology}

We have seen repeatedly in this paper that the Feynman diagram for the generalized Hall current can be expressed as a trace involving a unitary matrix $\xi$ and its derivatives, contracted with an epsilon tensor, where $\xi$ is the unitarized Euclidiean fermion operator in $d+1$ dimensions, $\xi =\tilde D/(\det \tilde D)^{2^{-k}}$ where $d+1=2k$ for even dimensions and $d=2k$ for odd dimensions. $\xi$ in general takes the form
\beq
\xi = a(p) + i \slashed{b}(p)\ ,\qquad a(p)^2 + b_\mu(p)  b_\mu(p) = 1\ .
\eqn{xidef}
\eeq
For an ordinary Dirac fermion, we have
\beq
a = \frac{m}{\sqrt{m^2+p^2}} \ ,\qquad b_\mu =  \frac{p_\mu}{\sqrt{m^2+p^2}}\ .
\eeq
We see that the $d+2$-component unit vector $\{a,b_\mu\}$
 represents a map from $(d+1)$-dimensional momentum space to the sphere $S^{d+1}$.  Note, however, that $\xi$  maps all of momentum space onto half of the sphere; as $p\to\infty$.  At $p=0$ we have $\xi=1$, while for infinite momentum,  $\xi \to i\hat p_\mu\gamma_\mu$ which can be thought of the equator of the sphere, where the poles are represented by the matrices $\pm 1$.  A map onto all of $S^{d+1}$ is described by $ U = \xi^2$, where for infinite momentum, $ U \to  -1$. Evidently the topology for a Dirac fermion in $(d+1)$-dimensional momentum space is described by the $O(d+1)$ nonlinear sigma model in $(d+1)$ dimensions.
 
 A convenient way to describe this  nonlinear sigma model is as an $SO(d+2)/SO(d+1)$ sigma model, which is what we do here. By computing $\sqrt{g}$, where $g_{\mu\nu}$ is the metric of the sigma model and $g$ is its determinant, we can simply express the volume form in terms of the matrix $\xi$, in a form easy to relate to the Feynman diagram calculations. 
 
 We define
\beq
 \xi &=&  e^{i\theta_\mu\gamma_\mu/4}= \cos\frac{\theta}{4} + i \hat{\slashed{\theta}} \sin\frac{\theta}{4} \ ,\cr
   U  &=& \xi^2 = e^{i\theta_\mu\gamma_\mu/2}  =  \cos\frac{\theta}{2} + i \hat{\slashed{\theta}} \sin\frac{\theta}{2} \ ,\cr
  A_a &=& \frac{i}{2}\left(\xi^\dagger\partial_a\xi - \xi\partial_a \xi^\dagger\right)  =  \CA_{a\mu} \gamma_\mu\ .
 \eqn{Udef}
  \eeq
For even $d+1$, it is easy to show that $ U $ parametrizes an $SO(d+2)/SO(d+1)$ sigma model.  We can write $ U $ as
 \beq
  U  = e^{i\theta_\mu \sigma'_{\mu,d+2}}\ ,\qquad \mu = 1,\ldots, d+1\ ,
 \eeq
 where 
 \beq
\sigma_{\mu\nu}' = \frac{i}{4} \left[\gamma_\mu',\gamma_\nu'\right]\ ,\qquad  \gamma_\mu' = i  \gamma_{d+2}\gamma_\mu\ ,\qquad \gamma_{d+2}' = \gamma _{d+2}\ , 
 \eeq
since  $\sigma'_{\mu,d+2} = \gamma_\mu/2$.    In this form we see that $ U (x)$ would describe the Goldstone bosons of the spontaneous symmetry breaking pattern $SO(d+2)\to SO(d+1)$, where the $\sigma_{\mu\nu}' $ generate  $SO(d+2)$, and the subset $\sigma'_{\mu,d+2}$ are the ``broken generators'' which are not also generators of the $SO(d+1)$ subalgebra.  A similar construction can be made for odd $d+1$.
 
 The metric for this sigma model is given by
 \beq
g_{\mu\nu} \propto \Tr \partial_\mu  U ^\dagger \partial_\nu  U  \propto   \Tr A_\mu A_\nu =\CA_{\mu\alpha} \CA_{\nu\beta}\Tr\gamma_\alpha\gamma_\beta \propto \left(\CA\CA^T\right)_{\mu\nu}\ ,
 \eeq
 where the proportionalities are all constant. Thus we have 
 \beq
  \sqrt{g} = N\det \CA\ .
  \eeq
  With the convention for both $SO(2k)$ and $SO(2k+1)$
  \beq
  \Tr \left[\gamma_{2k+1} \gamma_{\mu_1}\ldots\gamma_{\mu_{2k}}\right] = (2i)^{k} \epsilon_{\mu_1\ldots\mu_{2k}}\ ,
  \eqn{gamchi}
  \eeq
  we can rewrite this as
  \beq
  \sqrt{g} =N\epsilon_{\mu_1\ldots\mu_{d+1}} \times
   \begin{cases}
   \Tr \gamma_{d+2} A_{\mu_1} \ldots A_{\mu_{d+1}} & \text{even } d+1\\
  \Tr A_{\mu_1} \ldots A_{\mu_{d+1}}  & \text{odd } d+1
  \end{cases}\ ,
\eeq

With the normalization condition
\beq
\int d^{d+1}\theta\, \sqrt{g} = 1 .
\eqn{normV}
\eeq
 we have
\begin{equation}
\begin{aligned}
&d+1=2:&\   &\sqrt{g} &=&- \frac{i}{4\pi} \epsilon_{ij}\Tr \gamma_3 A_i A_j =\frac{i}{16\pi}\epsilon_{ij} \Tr[\gamma_\chi(\partial_i U)(U^\dagger\partial_j U)\\
&&&&=&
 \frac{1}{8\pi} \left(\frac{\sin\theta/2}{\theta} \right)\ ,\\  
&d+1=3:&\  &\sqrt{g} &=&\frac{i}{3\pi^2} \epsilon_{ijk}\Tr A_i A_j A_k =\frac{1}{24\pi^2} \epsilon_{ijk}\Tr(U^\dagger\partial_i U)(U^\dagger\partial_j U)(U^\dagger\partial_k U)\\
&&&&
=& \frac{1}{4\pi^2}\left( \frac{\sin\theta/2}{\theta} \right)^2\ ,\\ 
&d+1=4:&\   &\sqrt{g} &= &-\frac{1}{16 \pi^2} \epsilon_{ijk\ell}\Tr \gamma_5 A_i A_j A_k A_\ell =
-\frac{1}{256\pi^2}\epsilon_{ijk\ell}\Tr[\gamma_5(\partial_i U)(U^\dagger\partial_j U)(U^\dagger\partial_k U)(U^\dagger\partial_\ell U) \\
&&&&=& \frac{3}{16\pi^2}\left( \frac{\sin\theta/2}{\theta} \right)^3\ ,\\  
&d+1=5:&\   &\sqrt{g} &= &\frac{1}{15\pi^3}\epsilon_{ijk\ell m}\Tr A_i A_j A_k A_\ell A_m \\
&&&&=&
\frac{i}{480\pi^3}\epsilon_{ijk\ell m}\Tr[(U^\dagger \partial_i U)(U^\dagger\partial_j U)(U^\dagger\partial_k U)(U^\dagger\partial_\ell U)(U^\dagger\partial_m U) \\
&&&&=& i\frac{1}{2\pi^3}\left( \frac{\sin\theta/2}{\theta} \right)^4\ .\\  
\end{aligned}
\eqn{rtg}
\end{equation}

The integral in \eq{normV} has an interpretation other than as the normalized volume integral: with the definitions in \eq{xidef} and \eq{Udef} it is the winding number of a map $U(p)$ from momentum space to $SO(d+2)/SO(d+1)\cong S^{d+1}$.  Since $U(p)\to -1$ as $|\bfp|\to \infty$ in any direction, momentum space is effectively compactified to $S^{d+1}$ and so the winding number is an element of the homotopy group $\pi_n(S^n)$ with $n=d+1$. 
\section{Gapping the Majorana edge states with interactions in $1+1$ dimensions}
\label{sec:appendix_mod8}

Here we give a simple argument for a result of \cite{Fidkowski:2009dba}, that one can fully gap the edge states in the $1+1$ dimensional Majorana model when the number of flavors is $2N=0\mod 8$, and not for other numbers of flavors. For a pedagogical review, see \cite{bernevig2017topological}

We have seen in \S~\ref{sec:interactions} that for $2N$ flavors, the most general  time reversal invariant interactions can be  written as sums of totally antisymmetrized products of $4k$ $SO(2N)$ gamma matrices acting on the $2^N$-dimensional Hilbert space of degenerate ground states.  Call this interaction matrix $H_\text{int}$, and consider the effect on $H_\text{int}$ of two special matrices: $\gamma_{2N+1}$, which anti-commutes with all the other gamma matrices,   and $\wtC$ which has the following properties\footnote{Note that the conventional  definition of  the charge conjugation matrix $C$ satisfies $C \gamma_\mu C =-\gamma_\mu^T $ whereas $\wtC$ defined here satisfies that equation for $SO(4k+2)$ but $\wtC \gamma_\mu \wtC =+\gamma_\mu^T $ for $SO(4k)$; both   $C$  and  $\wtC$ to conjugate the generators, $\sigma_{\mu\nu}\to -\sigma_{\mu\nu}^T$, and the alternating sign better serves our purpose here.  For $SO(4k)$ the conventional $C$ is given by multiplying  $\wtC$   by the chiral matrix $\gamma_{4k+1}$. }:
\beq
\wtC &=& \wtC^\dagger = \wtC^{-1}\ , \cr &&\cr&&\cr
\wtC \gamma_\mu \wtC &=& \begin{cases} 
-\gamma_\mu^T & 2N=8k+2, 8k+6 \\ 
+\gamma_\mu^T & 2N=8k, 8k+4\ .
 \end{cases}\ ,
 \cr &&\cr&&\cr
  \wtC^T &=& 
\begin{cases} 
+ \wtC & 2N=8k,\,8k+6\\
-  \wtC & 2N=8k+2,\,8k+4\ .
\end{cases}\ ,\cr &&\cr&&\cr
  \wtC\gamma_{2N+1} \wtC &= &
\begin{cases} 
+\gamma_{2N+1} & 2N=4k,\\
-\gamma_{2N+1} & 2N=4k+2\ .
\end{cases}\ .
\eqn{Cprop}\eeq
 An explicit representation of $\wtC$ can be easily found using the following recursive definition for the gamma matrices of $SO(2N)$.  For $SO(2)$ we take the Pauli matrices as our gamma matrices:
\beq
 SO(2):\qquad &\gamma^{(2)}_i = \sigma_i\ ,\qquad i = 1,\ldots,3\ ,
 \eeq
and then for $SO(2N)$ for $N\ge 2$ we take
\beq
\gamma_i^{(2N)} &=& \sigma_1\otimes \gamma^{(2N-2)}_i\ ,\qquad  i = 1,\ldots, 2N-1\ ,\cr &&\cr
\gamma_{2N}^{(2N)} &=& \sigma_2\otimes 1\ ,\cr &&\cr
\gamma_{2N+1}^{(2N)} &=& \sigma_3\otimes 1\ .
\eeq
In this basis the matrix $\wtC$ is found to be a direct product of matrices alternating between $\sigma_2$ and $\sigma_3$:
\beq
\wtC^{(2)} = \sigma_2\ ,\ \wtC^{(4)} = \sigma_3\otimes \sigma_2\ ,\   \wtC^{(6)} = \sigma_2\otimes \sigma_3\otimes \sigma_2\ ,\ \wtC^{(8)} = \sigma_3\otimes \sigma_2\otimes \sigma_3\otimes \sigma_2\ , 
\eeq 
and so on. It can be easily verified in this basis that $\wtC$ possesses the properties in \eq{Cprop}.

Now it is easy to prove that the eigenvalues of $H_\text{int}$ have to be at least doubly degenerate for all $SO(2N)$ groups except $2N=8k$, where $k$ is an integer.   $H_\text{int}$ consists of sums of products of $4k$  $\gamma$-matrices totally antisymmetrized in their indices, with real coefficients, it follows that
\beq
H_\text{int}=H_\text{int}^\dagger\ ,\qquad [H_\text{int},\gamma_{2N+1}]=0\ ,\qquad   \wtC H_\text{int}   \wtC = H^T_\text{int}\ .
\eeq

Since $[H_\text{int},\gamma_{2N+1}]=0$ we can simultaneously diagonalize $H_\text{int}$ and $\gamma_{2N+1}$ with eigenvectors $\psi_{n,\sigma}$ where
\beq
H_\text{int}\psi_{n,\sigma}=\lambda_n\psi_{n,\sigma}\ ,\qquad \gamma_{2N+1}\psi_{n,\sigma} = \sigma \psi_{n,\sigma} \ ,\qquad \lambda_n\in {\mathbb R}\ ,\quad \sigma=\pm 1.
\eeq
Now consider their action on the vector $\chi_{n,\sigma} \equiv    \wtC\psi_{n,\sigma}^*$:
\beq
H_\text{int}  \chi_{n,\sigma} &=&   \wtC(  \wtC H_\text{int}  \wtC) \psi_{n,\sigma}^* = \wtC H_\text{int}^T \psi_n^* =  \wtC H_\text{int}^* \psi_n^* =\lambda_n  \wtC\psi_n^*=\lambda_n \chi_{n,\sigma} \ ,
\eeq
and
\beq
\gamma_{2N+1}\chi_{n,\sigma}  &=&  \wtC( \wtC\gamma_{2N+1}  \wtC) \psi_{n,\sigma}^* = \begin{cases}   \wtC \gamma_{2N+1} \psi_{n,\sigma}^* = +\sigma\chi_{n,\sigma} & 2N = 4k\\
-  \wtC \gamma_{2N+1} \psi_{n,\sigma}^* =- \sigma\chi_{n,\sigma} & 2N = 4k+2\ .
\end{cases}
\eeq
Thus $\chi_{n,\sigma} $ is an eigenstate of $H_\text{int}$ with eigenvalue $\lambda_n$, and now we would like to know if it is proportional to $ \psi_{n,\sigma}$, in which case we have learned nothing, or orthogonal to $ \psi_{n,\sigma}$, in which case we have shown that the eigenvalue $\lambda$ is at least doubly degenerate. 

First of all, we see that while $\gamma_{2N+1}\psi_{n,\sigma} = \sigma \psi_{n,\sigma} $ we have $\gamma_{2N+1}\chi_{n,\sigma} = -\sigma \chi_{n,\sigma} $ for $2N=4k'+2 = 8k+2,\,8k+6$, proving that $\psi_{n,\sigma} $ and $\chi_{n,\sigma} $ are indeed orthogonal in these cases and the spectrum is doubly degenerate.
 We can also directly compute their inner product and find
 \beq
 \chi_{n,\sigma}^\dagger\psi_{n,\sigma} = \psi_{n,\sigma}^T  \wtC\psi_{n,\sigma}
 \eeq
 which equals zero whenever $ \wtC$ is antisymmetric, which occurs for $2N= 8k+2, \,8k+4$ -- and so the spectrum is doubly degenerate in these cases also.
 
 Putting the two results together we see double degeneracy for $2N=8k+2,\,8k+4,\,8k+6$, leaving only $2N=8k$ as a possible candidate for $H_\text{int}$ to have unique eigenvalues.

Now we can ask: for $SO(8k)$ do we need to go beyond the $\gamma^4$ terms in $H_\text{int}$ in order to gap all of the edge states? For this we will just count parameters.  $H_\text{int}$ is $2^N$ dimensional, so we should be able to obtain $2^N$ different eigenvalues if we have at least $2^N$ parameters in $H_\text{int}$. (This is overkill, since we only need the lowest eigenvalue to be unique).  The number of independent antisymmetric 4-index tensors, whose indices can take $2N$ values is $2N!/[ 4!(2N-4)!]$ which is greater than $2^N$ for $N\ge 3$. Therefore a purely 4-fermion interaction can  gap all the fermions for every $SO(8k)$.  

A simple example  in the eight flavor model  of a time reversal invariant interaction that gaps all of the edge states is
\beq
H_\text{int} =\omega\left( \hat c_1  \hat c_2  \hat c_3  \hat c_4  + h.c. + 1\right)\ ,
\eqn{Hint8}
\eeq
where the $\hat c_i$ ladder operators were defined in \eq{cdef} and we assume $\omega>0$.  
(Under time reversal, $\hat c_i\leftrightarrow - \hat c_i^\dagger$). 
The eigenstates of $H_\text{int}$ include a unique groundstate with eigenvalue $E=0$, fourteen degenerate states with eigenvalue $E=\omega$, and a unique maximal state with eigenvalue $2\omega$.   The eigenstates corresponding to the minimum and maximum energy are linear combinations of the empty and fully occupied states, $(\ket{0000}\mp\ket{1111})/\sqrt{2}$, where $\hat c_i \ket{0000}=0$, and $\ket{1111}  = \hat c^\dagger_4 \hat c^\dagger_3 \hat c^\dagger_2 \hat c^\dagger_1 \ket{0000}$.                                                                                                                                                                                                                                                                                                                                                                                                                                                                                                                                                                                                                                                                                                                                                                                                                                                                                                                                         

The interaction in \eq{Hint8} can be realized in our $1+1$-dimension Lagrangian by the term 
\beq
\CL_\text{int} = \omega \left[ \left( \psi_1 + i \psi_5\right)^T C \gamma_\chi \left( \psi_2 + i \psi_6\right)\right]\left[ \left( \psi_3 + i \psi_7\right)^T C \gamma_\chi \left( \psi_4 + i \psi_8\right)\right] + h.c.\ .
\eqn{Lgap}
\eeq

\bibliography{index}

\begin{thebibliography}{10}

\bibitem{Ludwig_2015}
Andreas W~W Ludwig.
\newblock Topological phases: classification of topological insulators and
  superconductors of non-interacting fermions, and beyond.
\newblock {\em Physica Scripta}, T168:014001, dec 2015.

\bibitem{PhysRevB.85.085103}
Xiao-Gang Wen.
\newblock Symmetry-protected topological phases in noninteracting fermion
  systems.
\newblock {\em Phys. Rev. B}, 85:085103, Feb 2012.

\bibitem{Ryu_2010}
Shinsei Ryu, Andreas~P Schnyder, Akira Furusaki, and Andreas W~W Ludwig.
\newblock Topological insulators and superconductors: tenfold way and
  dimensional hierarchy.
\newblock {\em New Journal of Physics}, 12(6):065010, jun 2010.

\bibitem{Kitaev:2009mg}
Alexei Kitaev.
\newblock {Periodic table for topological insulators and superconductors}.
\newblock {\em AIP Conf. Proc.}, 1134(1):22--30, 2009.

\bibitem{PhysRevB.78.195125}
Andreas~P. Schnyder, Shinsei Ryu, Akira Furusaki, and Andreas W.~W. Ludwig.
\newblock Classification of topological insulators and superconductors in three
  spatial dimensions.
\newblock {\em Phys. Rev. B}, 78:195125, Nov 2008.

\bibitem{Callan:1984sa}
Curtis~G. Callan, Jr. and Jeffrey~A. Harvey.
\newblock {Anomalies and fermion zero modes on strings and domain walls}.
\newblock {\em Nucl. Phys. B}, 250:427--436, 1985.

\bibitem{kaplan2021index}
David~B Kaplan and Srimoyee Sen.
\newblock Index theorems, generalized hall currents, and topology for gapless
  defect fermions.
\newblock {\em Phys. Rev. Lett.}, 128(25):251601, 2022.

\bibitem{fujikawa1979path}
Kazuo Fujikawa.
\newblock Path-integral measure for gauge-invariant fermion theories.
\newblock {\em Phys. Rev. Lett.}, 42(18):1195, 1979.

\bibitem{Kaplan:1999jn}
David~B. Kaplan and Martin Schmaltz.
\newblock {Supersymmetric Yang-Mills theories from domain wall fermions}.
\newblock {\em Chin. J. Phys.}, 38:543--550, 2000.

\bibitem{kitaev2001unpaired}
A~Yu Kitaev.
\newblock Unpaired majorana fermions in quantum wires.
\newblock {\em Physics-uspekhi}, 44(10S):131, 2001.

\bibitem{Fidkowski:2009dba}
Lukasz Fidkowski and Alexei Kitaev.
\newblock {The effects of interactions on the topological classification of
  free fermion systems}.
\newblock {\em Phys. Rev. B}, 81:134509, 2010.

\bibitem{thouless1982quantized}
David~J Thouless, Mahito Kohmoto, M~Peter Nightingale, and Marcel den Nijs.
\newblock Quantized {H}all conductance in a two-dimensional periodic potential.
\newblock {\em Phys. Rev. Lett.}, 49(6):405, 1982.

\bibitem{niu1984quantised}
Qian Niu and DJ~Thouless.
\newblock Quantised adiabatic charge transport in the presence of substrate
  disorder and many-body interaction.
\newblock {\em Journal of Physics A: Mathematical and General}, 17(12):2453,
  1984.

\bibitem{niu1985quantized}
Qian Niu, D~J Thouless, and Yong-Shi Wu.
\newblock Quantized {H}all conductance as a topological invariant.
\newblock {\em Phys. Rev. B}, 31(6):3372, 1985.

\bibitem{wiki:periodic}
{Wikipedia contributors}.
\newblock Periodic table of topological invariants, 2018.
\newblock [Online; accessed 24-March-2022].

\bibitem{Jackiw:1975fn}
R.~Jackiw and C.~Rebbi.
\newblock {Solitons with fermion number 1/2}.
\newblock {\em Phys. Rev. D}, 13:3398--3409, 1976.

\bibitem{redlich1984gauge}
A~Norman Redlich.
\newblock Gauge noninvariance and parity nonconservation of three-dimensional
  fermions.
\newblock {\em Phys. Rev. Lett.}, 52(1):18, 1984.

\bibitem{ishikawa1984chiral}
K~Ishikawa.
\newblock Chiral anomaly and quantized {H}all effect.
\newblock {\em Phys. Rev. Lett.}, 53(17):1615, 1984.

\bibitem{Kaplan:1992bt}
David~B. Kaplan.
\newblock {A method for simulating chiral fermions on the lattice}.
\newblock {\em Phys. Lett. B}, 288:342--347, 1992.

\bibitem{Jansen:1992tw}
Karl Jansen and Martin Schmaltz.
\newblock {Critical momenta of lattice chiral fermions}.
\newblock {\em Phys. Lett. B}, 296:374--378, 1992.

\bibitem{Golterman:1992ub}
Maarten F.~L. Golterman, Karl Jansen, and David~B. Kaplan.
\newblock {Chern-Simons currents and chiral fermions on the lattice}.
\newblock {\em Phys. Lett. B}, 301:219--223, 1993.

\bibitem{ishikawa1987microscopic}
Kenzo Ishikawa and Toyoki Matsuyama.
\newblock A microscopic theory of the quantum {H}all effect.
\newblock {\em Nucl. Phys. B}, 280:523--548, 1987.

\bibitem{cheng2013fermion}
Bobby Cheng and Chris Ford.
\newblock Fermion zero modes for abelian {BPS} monopoles.
\newblock {\em Phys. Lett. B}, 720(1-3):262--264, 2013.

\bibitem{PhysRevB.85.245132}
Shinsei Ryu and Shou-Cheng Zhang.
\newblock Interacting topological phases and modular invariance.
\newblock {\em Phys. Rev. B}, 85:245132, Jun 2012.

\bibitem{PhysRevB.88.064507}
Hong Yao and Shinsei Ryu.
\newblock Interaction effect on topological classification of superconductors
  in two dimensions.
\newblock {\em Phys. Rev. B}, 88:064507, Aug 2013.

\bibitem{Gu:2013azn}
Zheng-Cheng Gu and Michael Levin.
\newblock {The effect of interactions on 2D fermionic symmetry-protected
  topological phases with Z2 symmetry}.
\newblock {\em Phys. Rev. B}, 89:201113, 2014.

\bibitem{PhysRevB.92.085114}
Tsuneya Yoshida and Akira Furusaki.
\newblock Correlation effects on topological crystalline insulators.
\newblock {\em Phys. Rev. B}, 92:085114, Aug 2015.

\bibitem{Gu:2015wli}
Yingfei Gu and Xiao-Liang Qi.
\newblock Axion field theory approach and the classification of interacting
  topological superconductors.
\newblock {\em arXiv:1512.04919}, 2015.

\bibitem{PhysRevB.90.245120}
Yi-Zhuang You and Cenke Xu.
\newblock Symmetry-protected topological states of interacting fermions and
  bosons.
\newblock {\em Phys. Rev. B}, 90:245120, Dec 2014.

\bibitem{PhysRevB.95.195108}
Xue-Yang Song and Andreas~P. Schnyder.
\newblock Interaction effects on the classification of crystalline topological
  insulators and superconductors.
\newblock {\em Phys. Rev. B}, 95:195108, May 2017.

\bibitem{PhysRevLett.117.206405}
Raquel Queiroz, Eslam Khalaf, and Ady Stern.
\newblock Dimensional hierarchy of fermionic interacting topological phases.
\newblock {\em Phys. Rev. Lett.}, 117:206405, Nov 2016.

\bibitem{Wang:2022ucy}
Juven Wang and Yi-Zhuang You.
\newblock Symmetric mass generation.
\newblock {\em Symmetry}, 14(7):1475, 2022.

\bibitem{Wang:2013yta}
Juven Wang and Xiao-Gang Wen.
\newblock Non-perturbative regularization of 1+ 1d anomaly-free chiral fermions
  and bosons: On the equivalence of anomaly matching conditions and boundary
  gapping rules.
\newblock {\em arXiv:1307.7480}, 2013.

\bibitem{You:2014vea}
Yi-Zhuang You and Cenke Xu.
\newblock {Interacting Topological Insulator and Emergent Grand Unified
  Theory}.
\newblock {\em Phys. Rev. B}, 91(12):125147, 2015.

\bibitem{Wang:2018ugf}
Juven Wang and Xiao-Gang Wen.
\newblock {A Solution to the 1+1D Gauged Chiral Fermion Problem}.
\newblock {\em Phys. Rev. D}, 99(11):111501, 7 2018.

\bibitem{Zeng:2022grc}
Meng Zeng, Zheng Zhu, Juven Wang, and Yi-Zhuang You.
\newblock {Symmetric Mass Generation in the 1+ 1 Dimensional Chiral Fermion
  3-4-5-0 Model}.
\newblock {\em Phys. Rev. Lett.}, 128(18):185301, 2022.

\bibitem{Tong:2021phe}
David Tong.
\newblock Comments on symmetric mass generation in 2d and 4d.
\newblock {\em JHEP}, 2022(7):1--28, 2022.

\bibitem{PhysRevX.11.011063}
Shlomo~S. Razamat and David Tong.
\newblock Gapped chiral fermions.
\newblock {\em Phys. Rev. X}, 11:011063, Mar 2021.

\bibitem{Goldstone:1981kk}
Jeffrey Goldstone and Frank Wilczek.
\newblock {Fractional Quantum Numbers on Solitons}.
\newblock {\em Phys. Rev. Lett.}, 47:986--989, 1981.

\bibitem{you2014topological}
Yi-Zhuang You, Zhong Wang, Jeremy Oon, and Cenke Xu.
\newblock Topological number and fermion green's function for strongly
  interacting topological superconductors.
\newblock {\em Phys. Rev. B}, 90(6):060502, 2014.

\bibitem{catterall2017novel}
Simon Catterall and David Schaich.
\newblock Novel phases in strongly coupled four-fermion theories.
\newblock {\em Phys. Rev. D}, 96(3):034506, 2017.

\bibitem{you2018symmetric}
Yi-Zhuang You, Yin-Chen He, Cenke Xu, and Ashvin Vishwanath.
\newblock Symmetric fermion mass generation as deconfined quantum criticality.
\newblock {\em Phys. Rev. X}, 8(1):011026, 2018.

\bibitem{xu2021green}
Yichen Xu and Cenke Xu.
\newblock Green's function zero and symmetric mass generation.
\newblock {\em arXiv:2103.15865}, 2021.

\bibitem{t1976symmetry}
Gerard 't~Hooft.
\newblock Symmetry breaking through {B}ell-{J}ackiw anomalies.
\newblock {\em Phys. Rev. Lett.}, 37(1):8--11, 1976.

\bibitem{bernevig2017topological}
Andrei Bernevig and Titus Neupert.
\newblock Topological superconductors and category theory.
\newblock {\em Lecture Notes of the Les Houches Summer School: Topological
  Aspects of Condensed Matter Physics}, pages 63--121, 2017.

\end{thebibliography}
\bibliographystyle{unsrt}

 \end{document}